\documentclass[reprint,superscriptaddress]{revtex4-1}

\usepackage{amsmath}
\usepackage{amsfonts}
\usepackage{amssymb}
\usepackage{graphicx}
\usepackage{moreverb}
\usepackage{url}
\usepackage{epstopdf}
\usepackage{color}
\usepackage{multirow}
\usepackage{subfigure}
\usepackage{xfrac}
\usepackage{float}
\usepackage[title,titletoc,toc]{appendix}

\begin{document}

\def\mean#1{\left< #1 \right>}

\title{Simulations of water nano-confined between corrugated planes}
\author{Jon~Zubeltzu}
\affiliation{CIC nanoGUNE, 20018 Donostia-San Sebasti\'{a}n, Spain}
\affiliation{Departamento e Instituto de F\'{i}sica de la Materia Condensada, Universidad Aut\'{o}noma de Madrid, E-28049 Madrid, Spain}
\author{Emilio~Artacho}
\affiliation{CIC nanoGUNE, 20018 Donostia-San Sebasti\'{a}n, Spain}
\affiliation{Theory of Condensed Matter, Cavendish Laboratory, University of Cambridge, Cambridge CB3 0HE, United Kingdom}
\affiliation{Basque Foundation for Science Ikerbasque, 48011 Bilbao, Spain}
\affiliation{Donostia International Physics Center, 20018 Donostia-San Sebasti\'{a}n, Spain}
\date{\today}

\begin{abstract}

Water confined to nanoscale widths in two dimensions between ideal planar walls has been the subject of ample study, aiming at understanding the intrinsic response of water to confinement, avoiding the consideration of the chemistry of actual confining materials. In this work, we study the response of such nanoconfined water to the imposition of a periodicity in the confinement by means of computer simulations, both using empirical potentials and from first-principles. For that we propose a periodic confining potential emulating the atomistic oscillation of the confining walls, which allows varying the lattice parameter and amplitude of the oscillation. We do it for a triangular lattice, with several values of the lattice parameter: one which is ideal for commensuration with layers of Ih ice, and other values that would correspond to more realistic substrates. For the former, the phase diagram shows an overall rise of the melting temperature. The liquid maintains a bi-layer triangular structure, however, despite the fact that it is not favoured by the external periodicity. The first-principles liquid is significantly affected by the modulation in its layering and stacking even at relatively small amplitudes of the confinement modulation. Beyond some critical modulation amplitude the hexatic phase present in flat confinement is replaced by a trilayer crystalline phase unlike any of the phases encountered for flat confinement. For more realistic lattice parameters, the liquid does not display higher tendency to freeze, but it clearly shows inhomogeneous behaviour as the strength of the rugosity increases. In spite of this expected inhomogeneity, the structural and dynamical response of the liquid is surprisingly insensitive to the external modulation. Although the first-principles calculations give a more triangular liquid than the one observed with empirical potentials (TIP4P/2005), both agree remarkably well for the main conclusions of the study.

\end{abstract}

\maketitle
\section{Introduction}

Although water has been one of the most studied substances in history, it still maintains a great scientific interest due to its complex behavior still defying understanding~\cite{ballh2o,mishima1998}. The study of water under strong confinement has attracted special attention during the last years~\cite{bertrand2013,kaneko2014,chen2015,corsetti2015b,bai2012,bai2003,koga1997,zubeltzu2016,corsetti2015a,zangi2003,zhao2014,zhu2015,giovambattista2009,zangi2004,chen2017,zubeltzu2016,han2010}. Its properties are strongly altered under these conditions, their understanding being important in biological, geological, and technological contexts~\cite{brovchenko2008,ball2003}. 

Computer simulations have been used to study the intrinsic behavior of nanoconfined water by means of simple confining potentials that depend on few parameters, avoiding possible effects that may be produced by particularities of the chosen confining substrate. Many new dynamical and structural properties of water have been observed, such as new crystalline phases~\cite{kaneko2014,chen2015,corsetti2015b,bai2012,bai2003,koga1997,zubeltzu2016,corsetti2015a,zangi2003,zhao2014,zhu2015,giovambattista2009,zangi2004,chen2017}, or unusual phase transitions~\cite{zubeltzu2016,han2010}. 

Han {\em et al.}~\cite{han2010} observed by classical molecular dynamics simulations using the TIP5P force-field model~\cite{mahoney2000} that bilayer water under planar confinement undergoes unusual freezing, which is continuous or discontinuous depending on where in the phase diagram. Bai and Zeng~\cite{bai2012} instead observe amorphous ice formation at similar conditions. Recent studies employing the TIP4P/2005 water model~\cite{abascal2005} combined with {\em ab initio} molecular dynamics simulations~\cite{corsetti2015b,zubeltzu2016} explain the origin of these observations proposing a scenario in which a continuous melting occurs, with the appearance of an intermediate hexatic phase following the Kosterlitz-Thouless-Halperin-Nelson-Young (KTHNY) theory for two-dimensional melting~\cite{kosterlitz1973,young1979,halperin1978,nelson1979}, and in which the hydrogen atoms are delocalized throughout the phase transition, while the oxygens freeze into a triangular ice. This phase, characterized by a large configurational entropy due to the large hydrogen disorder, can be quenched into amorphous ices, and undergoes a further discontinuous phase transition when the H atoms freeze, thereby accounting for the observations.

In order to obtain a deeper understanding of nanoconfined water, it is necessary to take into account the influence of the confining substrate. Computer simulations using atomistic surfaces~\cite{giovambattista2012,fitzner2016,al2017,kiselev2017,sosso2016} and realistic protein-like surfaces~\cite{reddy2010,khurana2006,cheng1998} have shown that the behavior of water changes drastically compared to the one observed using smooth surfaces. It has been observed in simulations with realistic surfaces that factors such as surface geometry, topography, and chemical heterogeneity play crucial roles~\cite{giovambattista2012,fitzner2016,mittal2010}. It is therefore important to study the effect of different aspects of the confining system.

 \begin{table*}[t!]
\caption{Values of various simulation parameters as defined in the text, that depend on the chosen lattice parameter.}
\centering
\begin{ruledtabular}
    \begin{tabular}{c | c @{\hspace{0.35 cm}}  c @{\hspace{0.35 cm}} c @{\hspace{0.35 cm}} c @{\hspace{0.35 cm}} c @{\hspace{0.35 cm}} c @{\hspace{0.35 cm}} c@{\hspace{0.35 cm}}  c }
    $a$~(\AA) & $\rho_{\text{LJ}}$~(\AA$^{-3}$) & $\epsilon_{12\text{-}6}$~(kcal/mol) & $l_{z_1}$~(\AA)  & $l_{z_2}$~(\AA) & $L_x^{\text{FFMD}}$~(\AA) & $L_y^{\text{FFMD}}$~(\AA) & $L_x^{\text{AIMD}}$~(\AA) & $L_y^{\text{AIMD}}$~(\AA) \\ \hline
    2.50 & 0.0905 & 0.0538 & -0.446  & -2.041 & 37.500 & 34.641 & 22.500 & 25.980 \\ 
    2.75 & 0.0680 & 0.0716 & -0.456  & -2.245 & 35.750 & 33.341 & 24.750 & 23.816 \\
    3.00 & 0.0524 & 0.0929 & -0.464  & -2.449 & 36.000 & 36.373 & 24.000 & 25.980 \\
    4.78 & 0.0129 & 0.3757 & -0.487  & -3.903 & 33.460 & 33.117 & 23.900 & 24.838 \\
    \end{tabular}
 \end{ruledtabular}
    \label{tabla1}
    \end{table*}

In this work we smoothly introduce a modulation to the planar potential employed in~\cite{han2010,zubeltzu2016}. This has been done before with large modulation periodicities~\cite{mittal2010} and in a different context (to study the ice nucleation on mineral surfaces of relevance for cloud formation~\cite{fitzner2016,cox2015a,cox2015b}). We study the effects produced by the topography of the confining substrate into the liquid, hexatic and solid phases described in~\cite{zubeltzu2016}. We do it by computational simulations using both, empirical potentials and first-principles calculations. For the confining walls, we propose a model based on a referential Lennard-Jones 9-3 potential plus Lennard-Jones explicit particles on its surface, allowing us to smoothly add a roughness into the confining potential and control the lattice parameter and energetic amplitude of the oscillation. We consider a triangular lattice: firstly, with a lattice parameter that commensurates with the honeycomb ice previously observed~\cite{zubeltzu2016}, and secondly, three more realistic values of the lattice parameter. For the former, although the honeycomb ice stabilizes as expected, the triangular structure disfavoured by the external modulation is maintained in the liquid. The activation of the external modulation has substantial effects on the first-principles liquid, affecting the stacking and the well-defined layering observed in the planar confinement. At high temperatures and densities, the frustration produced on the hexatic phase by the external modulation stabilizes a different tri-layer ice beyond a critical amplitude of the modulation. For more realistic lattice parameters, a clear dynamical and structural inhomogeneous behavior is observed in the liquid density, although other properties are much less affected.

\section{Methods}

We carry out computational simulations based on molecular dynamics with force-fields (FFMD), and {\em ab initio} molecular dynamics (AIMD). For the former, we use the LAMMPS code~\cite{plimpton1995} and the TIP4P/2005 force field to model the interaction among the water molecules, where the cutoff of the Lennard-Jones interaction is set to 12~$\text{\AA}$ and the particle-particle particle-mesh (PPPM) method~\cite{hockney1988} is used to compute the long-range Coulombic interactions. The rigidity of the molecules is treated using the SHAKE algorithm~\cite{ryckaert1977}. The timestep is set to 1 fs. Starting from a randomly set configuration of positions and velocities, during the first 60 ns of each FFMD run we adopt the constant particle number, volume, and temperature ensemble ({\em NVT}) and use the Nose-Hoover thermostat in order to control the temperature of the system. Then, 2 ns of dynamics are collected for data analysis. The dimensions of the rectangular cell ($L_x^{\text{FFMD}}\times L_y^{\text{FFMD}}$) are different for each lattice parameter in order to keep the periodicity of the cell (see Table~\ref{tabla1}). For system size testing, see~\cite{corsetti2015b}. 

For the AIMD calculations based on density functional theory, we use the SIESTA code~\cite{soler2002} with the vdW-DF functional with a fully non local correlation~\cite{dion2004}, devised to describe the van der Waals interactions, and PBE~\cite{perdew1996} exchange, especially tested for water~\cite{wang2011} (as in refs~\cite{corsetti2015b,zubeltzu2016}). Previous studies with this functional have shown noticeable improvements of the calculated radial distribution functions of water due to a better description of H bonds and reproduce the maximum of the diffusion with respect to the density~\cite{corsetti2013}.

The final configuration obtained from the FFMD calculations is annealed for 5 ps and then the {\em NVE} ensemble is used for at least 10 ps in AIMD while data are collected. The timestep is set to 0.5 fs. Due to the larger computational cost of such calculations, we reduce the size of the cell. Table~\ref{tabla1} shows the dimensions of the cell ($L_x^{\text{AIMD}}\times L_y^{\text{AIMD}}$) for each lattice parameter.

The model we propose to confine water to a thin film is a linear combination of a surface made of Lennard-Jones particles and a Lennard-Jones 9-3 potential. The latter is obtained by integrating semi-infinitely a piece of matter made of Lennard-Jones particles considering a density $\rho_{\text{LJ}}$~\cite{han2010,zubeltzu2016}. The proposed confining potential is:

\begin{equation}
\begin{split}
U(x,y,z)&=U_{9\text{-}3}(z) + \alpha\big\{U_{12\text{-}6}(x,y,z-l_{z_1})\\
&\quad-[U_{9\text{-}3}(z)-U_{9\text{-}3}(z-l_{z_2})]\big\},
\end{split}
\end{equation}
where:
 \begin{align} 
 U_{9\text{-}3}(z) & = 4\epsilon_{9\text{-}3}\Big[\Big(\dfrac{\sigma_{9\text{-}3}}{z}\Big)^9-\Big(\dfrac{\sigma_{9\text{-}3}}{z}\Big)^3\Big],\\ 
U_{12\text{-}6}(x,y,z) & = \sum_i^N4\epsilon_{12\text{-}6}\Big[\Big(\dfrac{\sigma_{12\text{-}6}}{\left|\mathbf{r}_i\right|}\Big)^{12}-\Big(\dfrac{\sigma_{12\text{-}6}}{\left|\mathbf{r}_i\right|}\Big)^6\Big],\\
\begin{split}
\left|\mathbf{r}_i\right|&=[(x-x_i)^2\\&\quad+(y-y_i)^2+(z-z_i)^2]^{1/2}.
\end{split}
\end{align}

On the right hand side of Eq.~(1) we add to the Lennard-Jones 9-3 potential a layer of Lennard Jones particles located at $z=l_{z_1}$, which introduces roughness into the potential, and remove the integrated piece of matter that represents this layer of Lennard-Jones particles $\big[U_{9\text{-}3}(z)-U_{9\text{-}3}(z-l_{z_2})\big]$. $l_{z_2}$ is the distance in $z$ between neighbor layers in the piece of matter made of Lennard-Jones particles. The second term on the right is multiplied by a parameter $\alpha$ which controls the strength of the roughness in the potential. Eq.~(1) can be rewritten simply:
\begin{equation}
\begin{split}
U(x,y,z)&=(1-\alpha)U_{9\text{-}3}(z)\\
&\quad+\alpha[U_{12\text{-}6}(x,y,z-l_{z_1})+U_{9\text{-}3}(z-l_{z_2})].
\end{split}
\end{equation}

\begin{figure*}[t!]
\centering
\subfigure[\ ]{\includegraphics[width=0.28\textwidth]{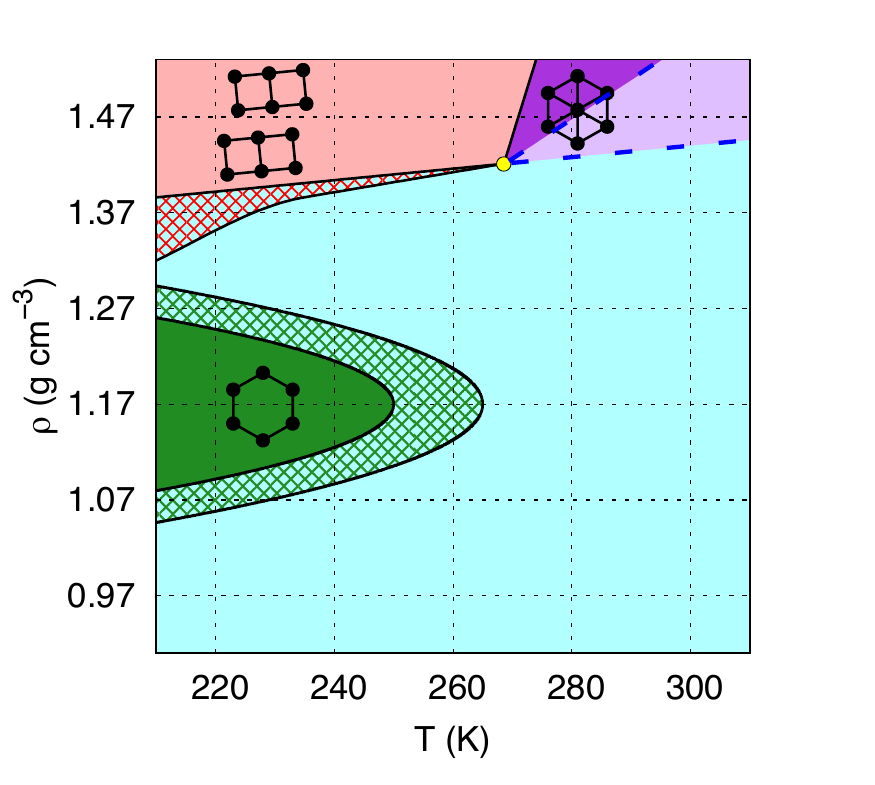}}
\subfigure[\ ]{\includegraphics[width=0.28\textwidth]{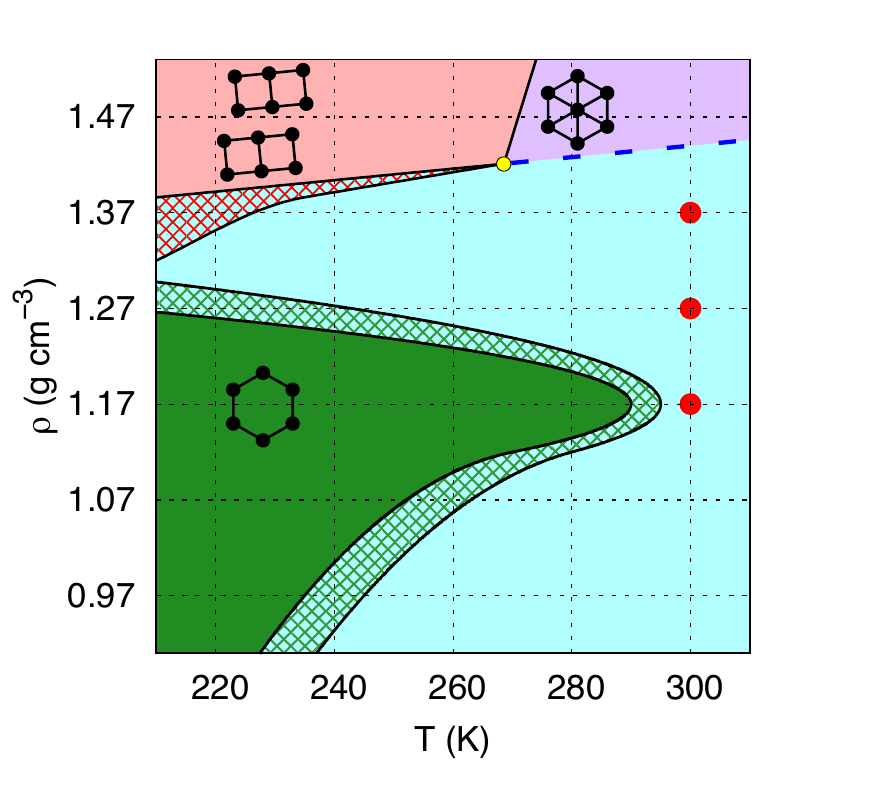}}
\subfigure[\ ]{\includegraphics[width=0.28\textwidth]{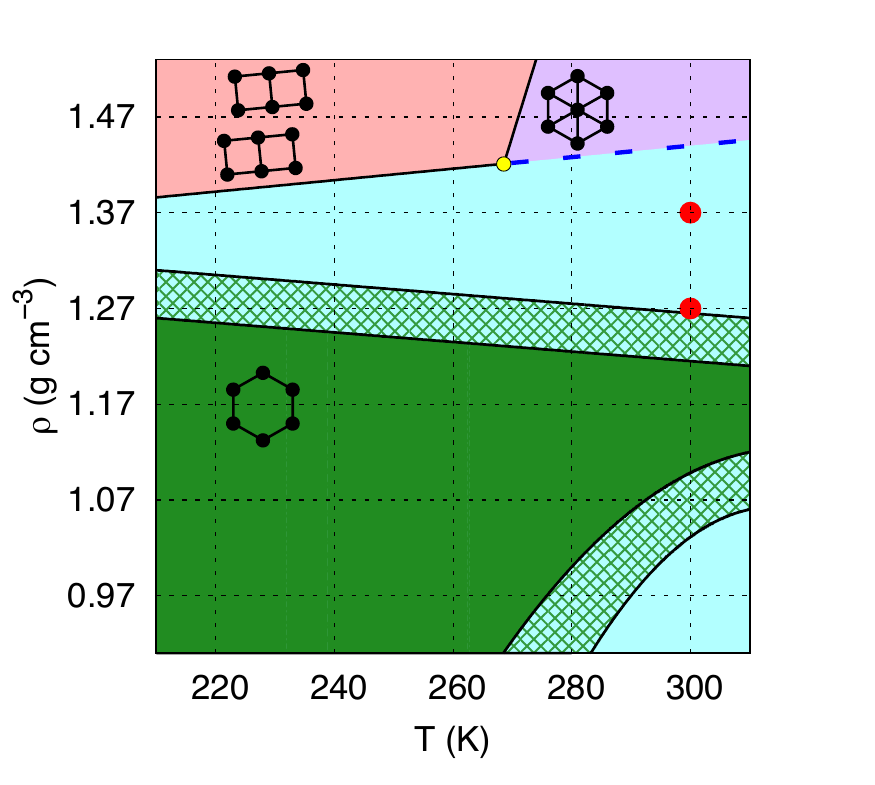}}
\subfigure[\ ]{\includegraphics[width=0.28\textwidth]{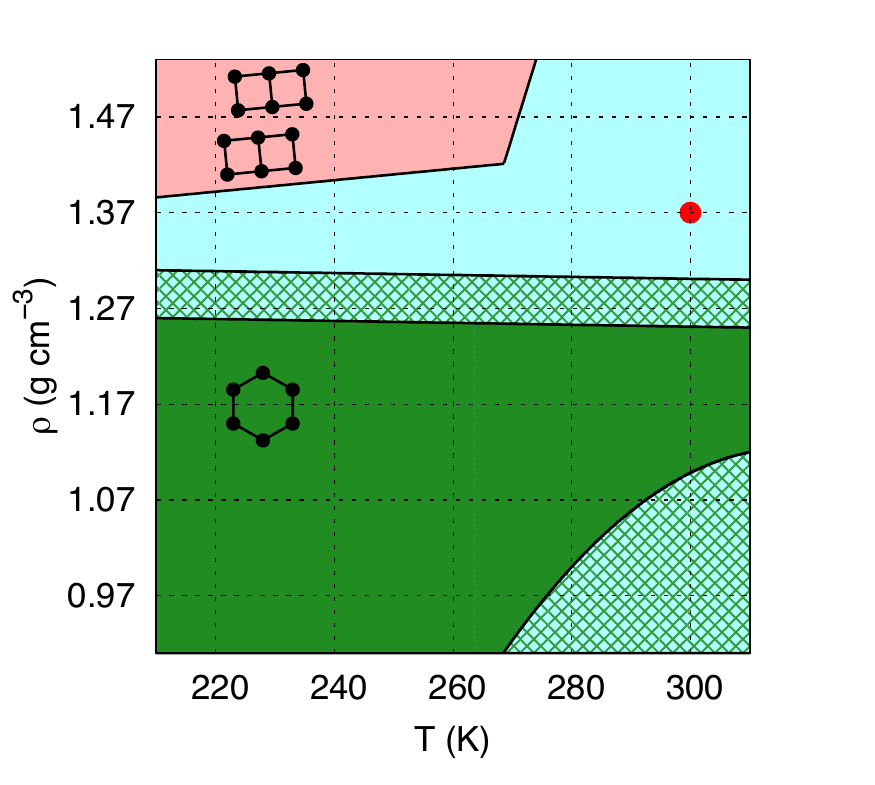}}
\subfigure[\ ]{\includegraphics[width=0.28\textwidth]{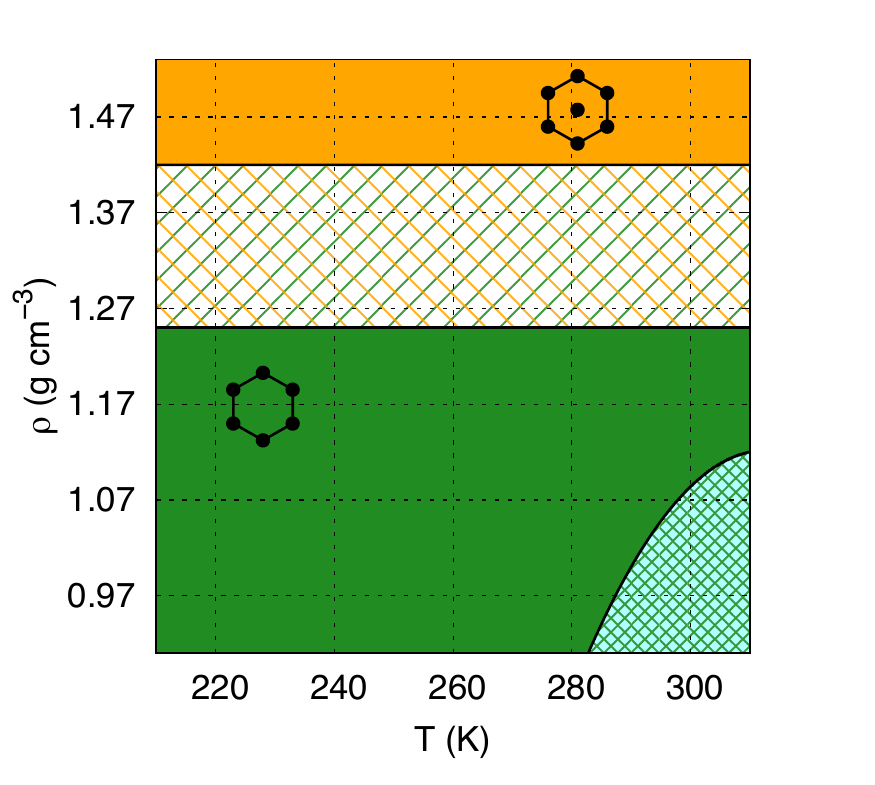}}
\subfigure[\ ]{\includegraphics[width=0.28\textwidth]{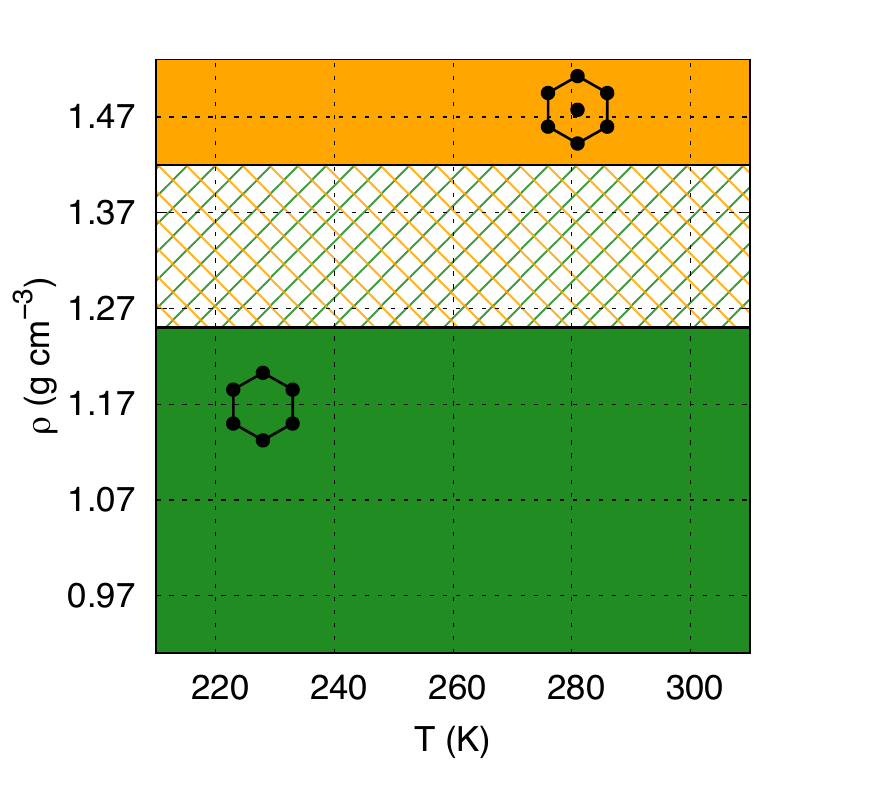}}
\caption{Summary of calculations and phase diagram constructed from the results for a lattice parameter $a=4.78$ $\text{\AA}$ and $\alpha=$ (a) 0.0, (b) 0.2, (c) 0.4, (d) 0.6, (e) 0.8, and (f) 1.0. The density is defined employing the effective distance $L^{\prime}_z$ (see text) as in Ref.~\cite{kumar2005}. The crossing points of the thin dashed grid are the points on the phase diagram sampled by FFMD, while the six red bullers show the points also calculated by AIMD. In the areas filled by a crosshatch we observe solid-liquid or solid-solid coexistence, consistent with having a first-order phase transition in an {\em NVT} ensemble. The black lines represent the first-order transition lines, while the blue dashed lines delimit the continuous phase transition lines among the liquid, hexatic and triangular ice solid phases. These transition lines were drawn semi quantitatively from the results obtained at the sampled points. Blue indicates the liquid phase. The other phases are described in the text.}
\label{fig:PD}
\end{figure*}

\begin{table}[t!]
\caption{Lowest energy modulation $E_b$ (see text) for a given value of $a$ and $\alpha$.}
\centering
\begin{ruledtabular}
    \begin{tabular}{ c  |  c  c  c  c }
    $\alpha$ & $ E_b^{a=2.50 \text{\AA}}$ (K) & $ E_b^{a=2.75 \text{\AA}}$ (K) & $ E_b^{a=3.00 \text{\AA}}$ (K) & $ E_b^{a=4.78 \text{\AA}}$ (K) \\ \hline
    0.2 & 3.88 & 6.93 & 11.05 & 44.38\\
    0.4 & 7.64 & 13.55 & 21.5 & 85.75\\
    0.6 & 11.32 & 20.01 & 31.98 & 136.61\\
    0.8 & 14.95 & 26.41 & 42.47 & 207.66\\
    1.0 & 18.54 & 32.79 & 53.16 & 342.19\\
    \end{tabular}
\end{ruledtabular}
\label{tabla2}
\end{table}

In our case, the $U_{9\text{-}3}$ potential only interacts with the oxygen atoms, and its parameters are chosen to mimic the interaction of water with solid paraffin as proposed in~\cite{lee1984} and used in~\cite{han2010,zubeltzu2016}: $\epsilon_{9\text{-}3}=0.2985$ kcal/mol and $\sigma_{9\text{-}3}=2.5$~\AA. The position of the confining potential is set so that the origin of the  $U_{9\text{-}3}$ potential is at $z$ = 0 $\text{\AA}$, and its symmetric potential at $z$ = 8 $\text{\AA}$, as in~\cite{han2010,zubeltzu2016}. The two layers of Lennard-Jones on each confining side are stacked in such a way that each atom in one layer has another one directly above (below) it, a disposition we will call AA stacking. From the condition that the piece of matter made of Lennard-Jones particles of density $\rho_{\text{LJ}}$ integrates into the Lennard-Jones 9-3 potential, we can obtain the relationship between the parameters of each potential:

 \begin{align}
 \sigma_{12\text{-}6}&=\sigma_{9\text{-}3}\left(\dfrac{15}{2}\right)^{1/6},\label{eq.6}\\
 \epsilon_{12\text{-}6}&=\sqrt{\dfrac{2}{15}}\dfrac{6\epsilon_{9\text{-}3}}{\rho_{\text{LJ}}\pi\sigma_{12\text{-}6}^3}.\label{eq.7}
 \end{align}
From Eq.~(\ref{eq.6}) we obtain $\sigma_{12\text{-}6}=3.498$~\AA, which is independent of the lattice parameter. The values of $ \epsilon_{12\text{-}6}$ for each lattice parameter are shown in Table~\ref{tabla1}.

During the integration, the Lennard-Jones particles are considered to be arranged into a fcc crystallographic configuration, where the surface is oriented in the $(111)$ direction conforming a triangular lattice. Given a triangular lattice parameter $a$, we deduce:

\begin{align}
 l_{z_2}&=a\sqrt{\frac{2}{3}},\\
 \rho_{\text{LJ}}&=\frac{\sqrt{2}}{a^3}.
 \end{align}

The position of the layer of Lennard-Jones particles $z=l_{z_1}$, is calculated so that the position of the minimum in $z$ of the mean confining potential $\mean{U}_{xy}(z)$ is independent of $\alpha$ (see Appendix~\ref{A1}). There here proposed confining potential requires a redefinition of the confinement width $L_z$, which is needed to calculate some physical magnitudes below. We use the one proposed in~\cite{kumar2005}:
\begin{equation} 
L^{\prime}_z=L_z-\dfrac{\sigma_{\text{9-3}}+\sigma_{\text{O}}}{2},
\end{equation}
where $\sigma_{\text{O}}$ = 3.1589 ~\AA~is the $\sigma$ parameter set for the Lennard-Jones interaction between the oxygen atoms in the TIP4P/2005 water model~\cite{abascal2005}. This gives $L^{\prime}_z=0.515$ nm. Table~\ref{tabla1} shows the values of the parameters used in the simulations for each lattice parameter. The value of $\alpha$ can be associated to an energy scale of relevance to the problem. Table~\ref{tabla2} shows the minima at on-top $xy$ positions (on top of Lennard-Jones particles) and the absolute minima of the potential for each lattice parameter and value of $\alpha$.

To obtain dynamical information of the simulations, we calculate the mean square displacement (MSD), defined as:
\begin{equation}
\Delta_{\text{xy}}(t)=[r(t)-r(0)]^2,
\end{equation} 
where, $r(t)=[x(t)^2+y(t)^2]^{1/2}$ is the position of the particle at time $t$ in the $xy$ plane. We then average this function over all the oxygens and different initial time steps:
\begin{equation}
\left<\Delta_{\text{xy}}(t)\right>=\dfrac{1}{N_{\text{cell}}t_{\text{steps}}}\sum_{\tau=0}^{t_{\text{steps}}}\sum_{i=1}^{N_{\text{cell}}}[r_i(t+\tau)-r_i(\tau)]^2.
\end{equation} 
The diffusivity $D$ is obtained from the Einstein's relation for a two-dimensional system:
\begin{equation}
\Delta_{\text{xy}}(t)=4Dt.
\label{diffusion}
\end{equation} 

The oxygen first-neighbor correlation function $C_{\text{O-O}}$ gives the proportion of initial in-plane nearest-neighbors of any particle that remain after time $t$. The nearest-neighborhood is defined by a circle of radius $r_0$ obtained from calculating the distance at which the radial distribution function shows a minimum between the first and second-neighbor peaks. Considering the function $n(t)$ that gives the number of nearest-neighbors that remain after time $t$, the averaged first-neighbor correlation function is defined as:

\begin{equation}
\left<C_{\text{O-O}}(t)\right>=\dfrac{1}{N_{\text{cell}}t_{\text{steps}}}\sum_{\tau=0}^{t_{\text{steps}}}\sum_{i=1}^{N_{\text{cell}}}\dfrac{n_i(t+\tau)}{n_i(\tau)}.
\end{equation}

The lateral pressure $P_{\parallel}$ and perpendicular pressure $P_\bot$ are calculated using the virial expression for the $x$ and $y$ directions, and calculating the force produced by oxygens on the wall, respectively, as in~\cite{kumar2005}. In order to obtain the density at which $P_{\parallel}=0$ keeping the lattice parameter constant (and hence, the dimensions of the cell), we carry out FFMD simulations on the $NVT$ ensemble with all the possible number of particles within a range such that the $P_{\parallel}=0$ point is known to be crossed. We then choose the $N$ that gives the closest value to $P_{\parallel}=0$.


\section{Results and Discussion}

We first discuss the results obtained with a triangular lattice parameter that is commensurated with honeycomb two-dimensional ice~\cite{zubeltzu2016}, and then we consider several values of $a$ in the range of what would be found with realistic confining materials, such as (111) faces of various metals. 

\subsection{Ideal commensuration}

\begin{figure}[b]
\centering
\subfigure[\ ]{\includegraphics[width=0.45\textwidth]{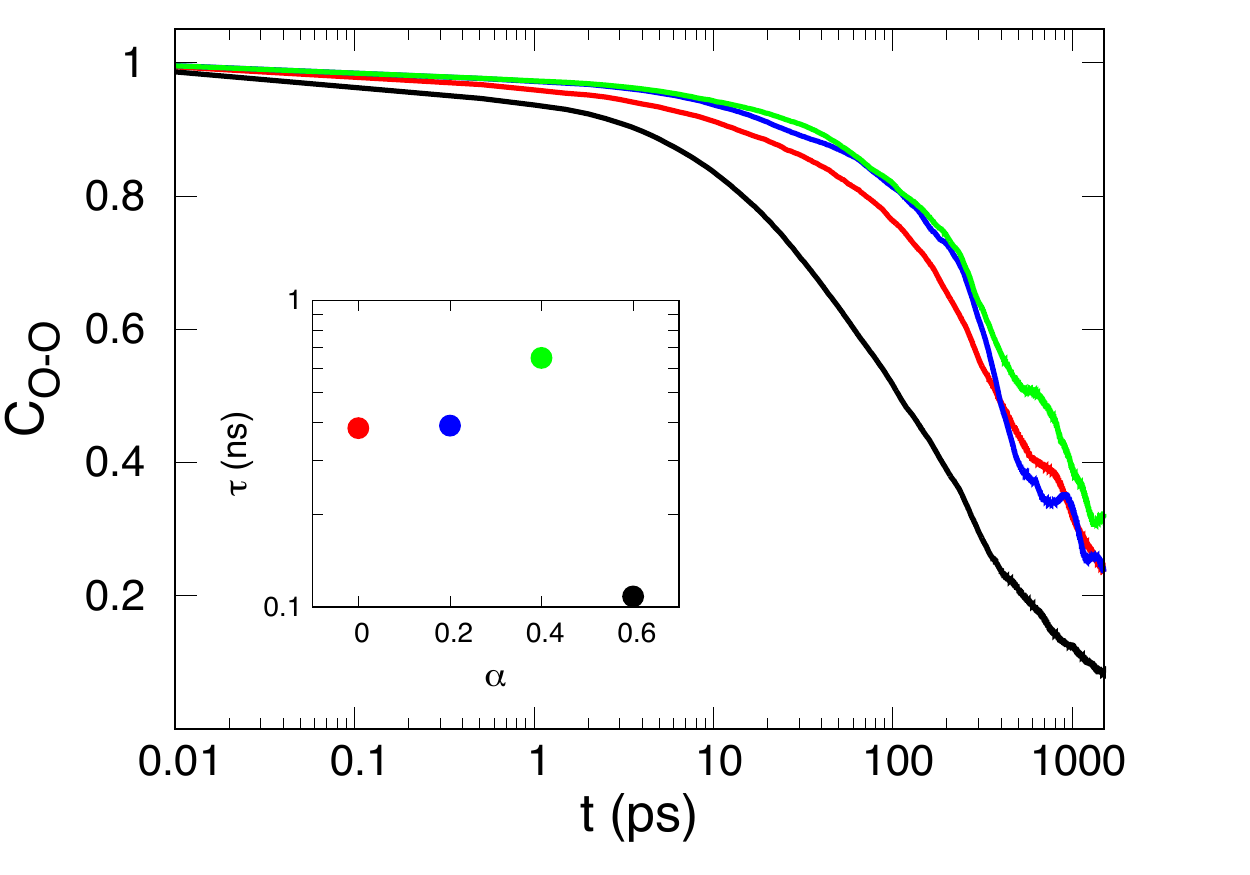}}\\
\subfigure[\ ]{\includegraphics[width=0.22\textwidth]{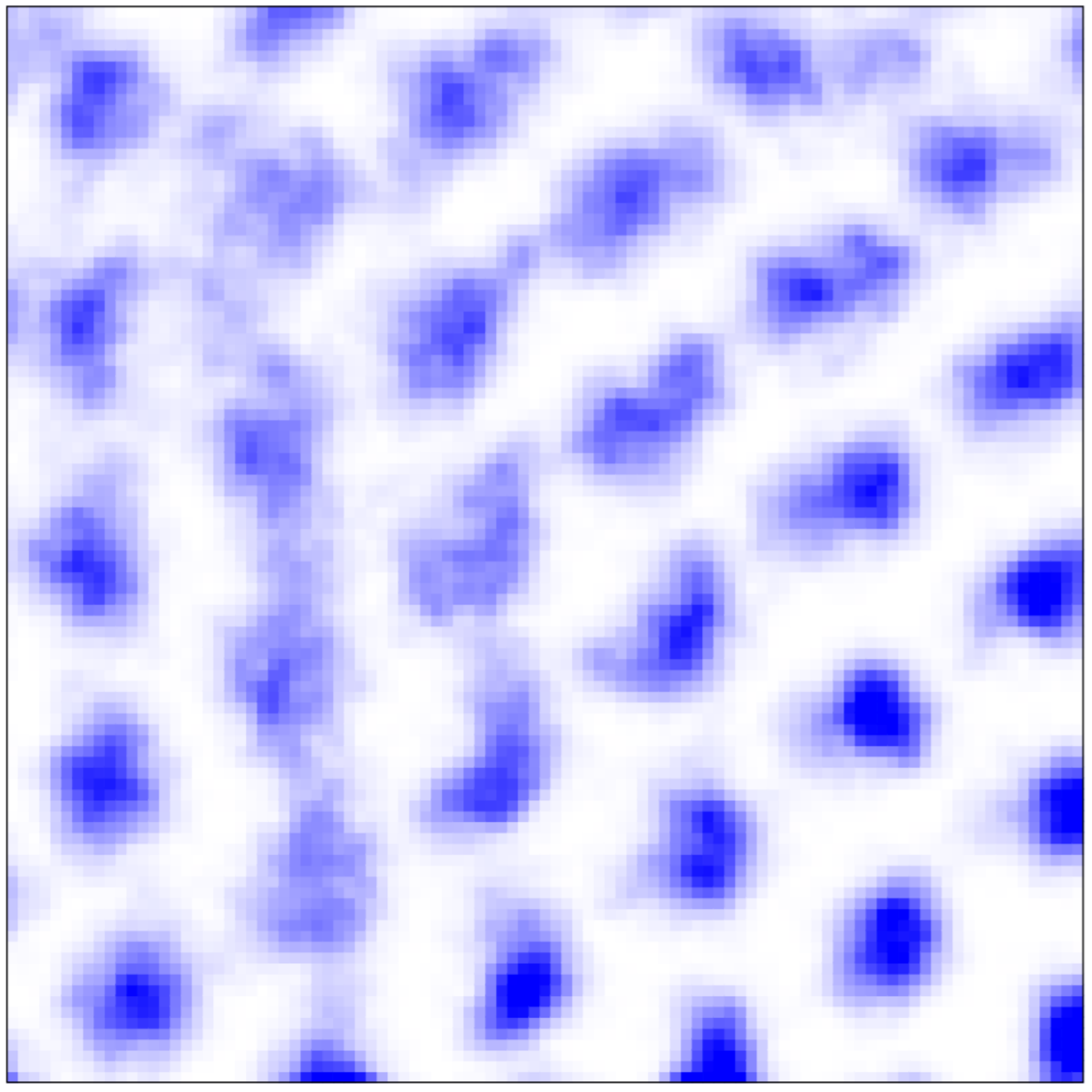}}
\subfigure[\ ]{\includegraphics[width=0.22\textwidth]{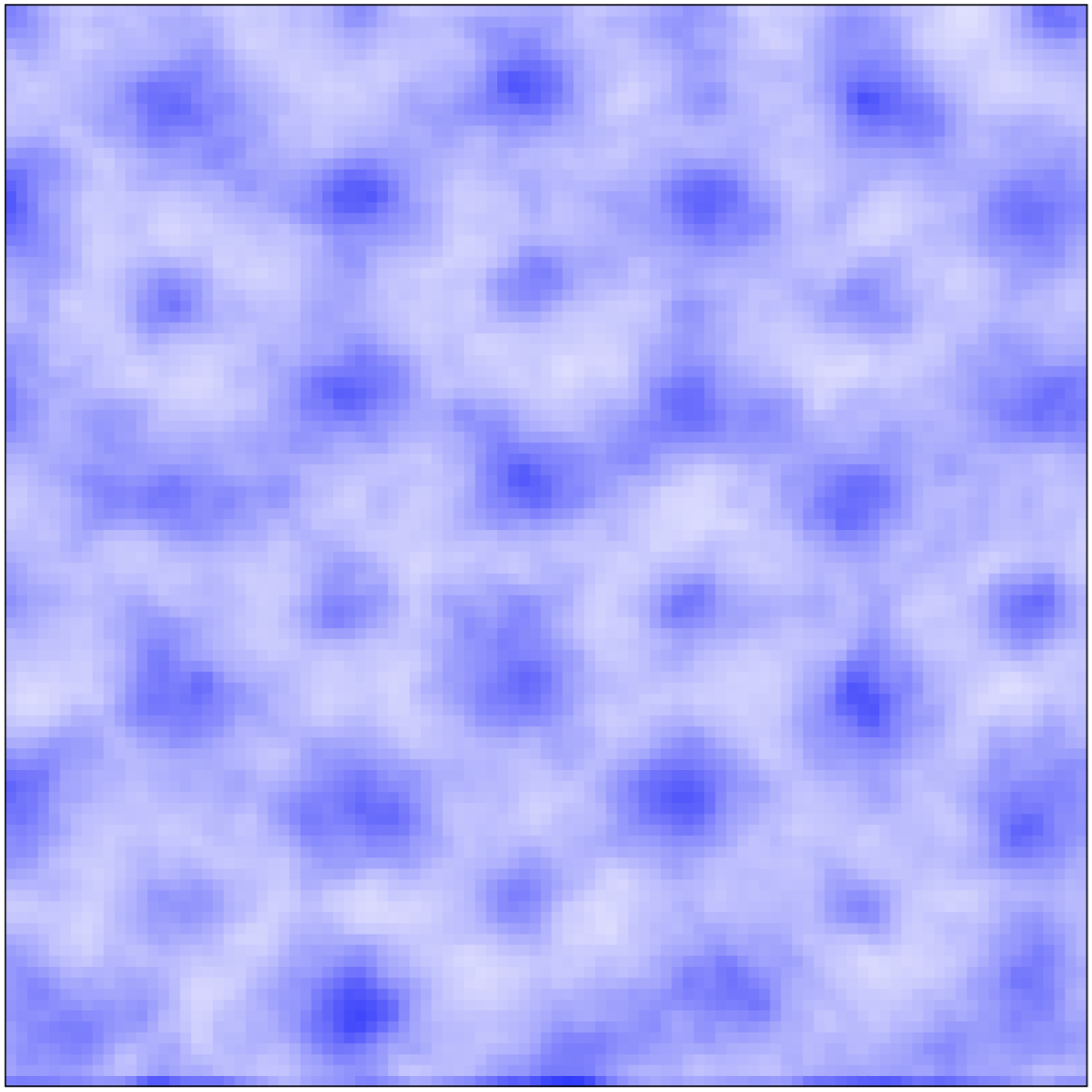}}
\caption{(a) Oxygen-oxygen first-neighbor correlation function at $T$ = 300 K and $\rho$ = 1.47 g~cm$^{-3}$ for $a$ = 4.78 $\text{\AA}$ and $\alpha$ = 0 (red), 0.2 (blue), 0.4 (green), and 0.6 (black). The inset shows the characteristic time for each curve. (b) and (c): averaged position of oxygens (blue) during 1 ns at $T$ = 300 K, $\rho$ = 1.47 g~cm$^{-3}$, and $\alpha$ = 0.4 (b), 0.6 (c). The images were obtained by averaging in time, within a window of dimensions $\left ( 15 \times 15 \right)$~\AA$^2$ in the simulation cell.}
\label{fig:cor}
\end{figure}

Taking into account that the oxygen-oxygen radial distribution function (RDF) of honeycomb ice shows an oxygen-oxygen $xy$ first-neighbor distance $r_{\text{O-O}}$ = 2.76 $\text{\AA}$, we choose a triangular lattice with a lattice parameter $a$ = $\sqrt{3}~r_{\text{O-O}}$ = 4.78 $\text{\AA}$. This produces an energy surface where the minima are positioned in a honeycomb structure with the distance between the nearest energy minima equal to $r_{\text{O-O}}$. Therefore, it should be ideal {\em a priori} for the establishment of a honeycomb ice monolayer, but it would disfavour the formation of triangular ice (one third of the oxygens would have to sit on maxima of the modulation potential). We study the structural and dynamical properties of water at different densities, temperatures, and values of $\alpha$.

\subsubsection{Phase diagram}

Fig.~\ref{fig:PD} shows the temperature-density phase diagrams for the system for $\alpha$ = 0.0, 0.2, 0.4, 0.6, 0.8, and 1.0, obtained from the FFMD calculations. The phases appearing in the diagram for $\alpha$ = 0.0 have been previously described in~\cite{zubeltzu2016}. In order to determine the phase at each calculated point, we have used four different indicators: radial distribution function, oxygen diffusion, and the oxygen $xy$ density histogram, the three of them in the {\em xy} plane, and the density profile of oxygen atoms along the confining direction, $z$. The $xy$ density histogram is obtained by averaging the $xy$ coordinates of the particles over 1 ns.

At $\rho\leq$ 1.27 g~cm$^{-3}$, the effect of increasing $\alpha$ is the expected one: the corrugation of the potential favours the structure of honeycomb ice and the melting point rapidly increases with $\alpha$. The observation of phase coexistence of honeycomb ice and liquid at different values of $\alpha$ indicates that the phase transition that connects these two phases keeps being first-order independently from the amplitude of the modulation (see Appendix~\ref{A2}). 

\begin{figure}[t]
\centering
\subfigure[\ ]{\includegraphics[width=0.3\textwidth]{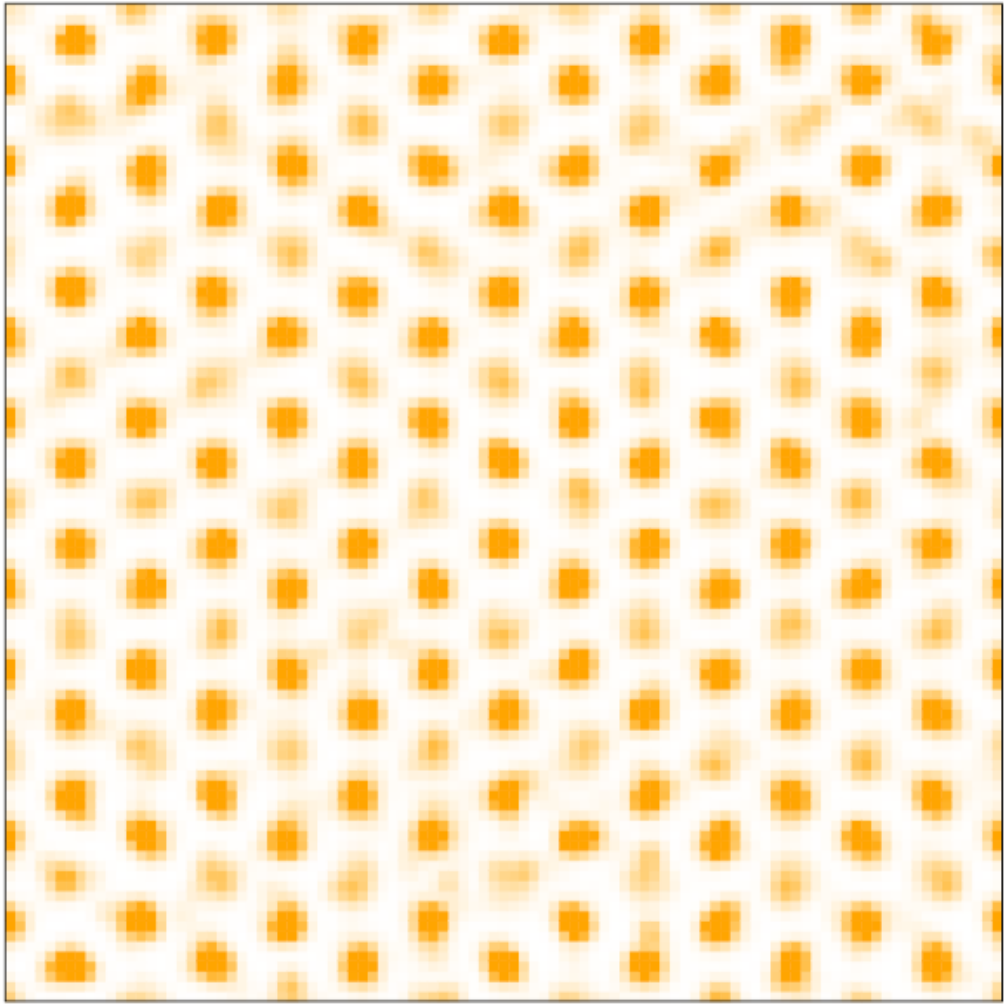}}\\
\subfigure[\ ]{\includegraphics[width=0.4\textwidth]{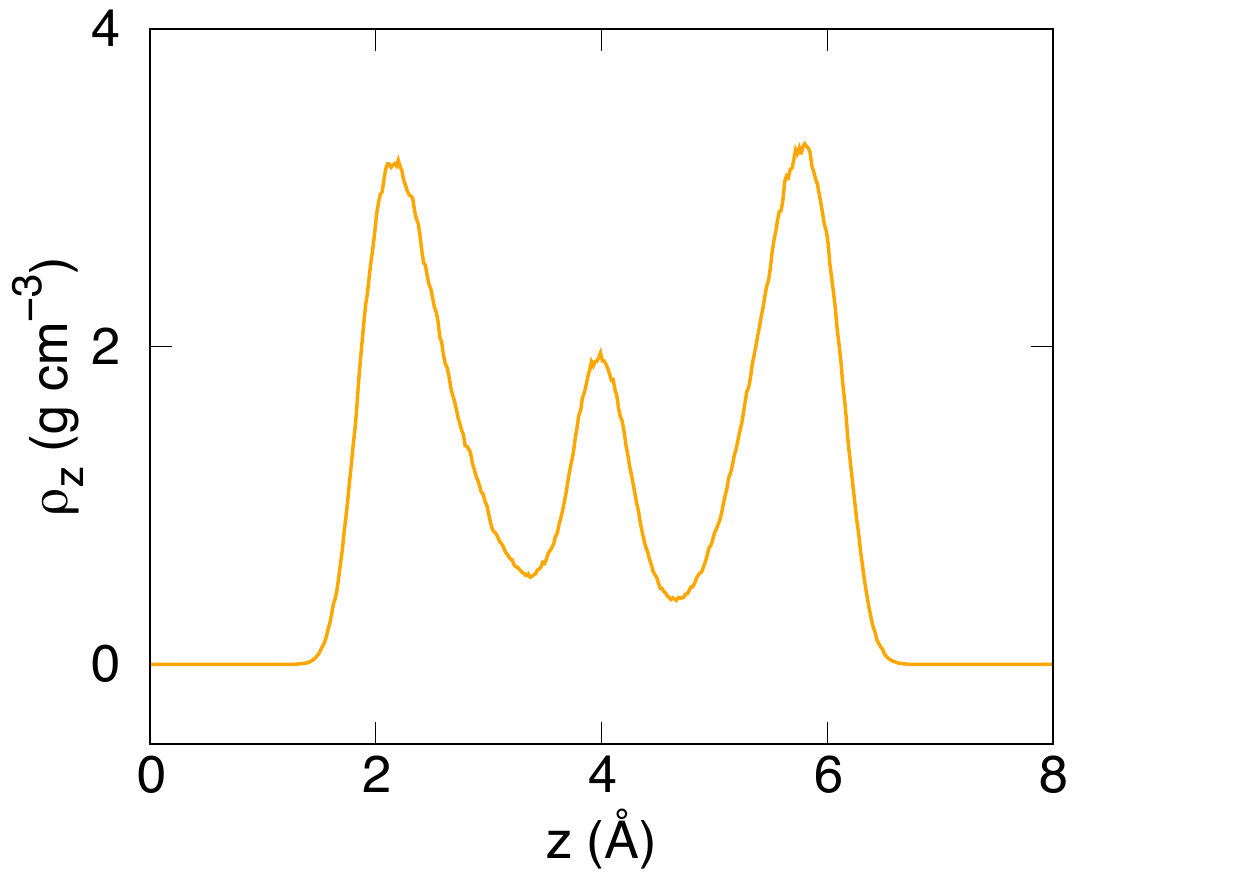}}
\caption{(a) $xy$ density histogram of oxygen atoms averaged over 1 ns, and (b) $z$ density profile of the oxygens, both at $T=260$ K, and $\rho$=1.47 g~cm$^{-3}$. The two exterior oxygen layers are arranged into a bilayer honeycomb lattice in AA stacking. The oxygens from the middle layer fill the centre of the honeycomb hexagons, coinciding with the $xy$ positions of the Lennard-Jones particles of the confining potential.}
\label{fig:ice}
\end{figure}

When $\alpha\leq0.6$, at $\rho$ = 1.37 g~cm$^{-3}$ water remains being liquid. At $\rho$ = 1.47 g~cm$^{-3}$ and low temperatures, the square-tubes ice observed for flat confinement~\cite{zubeltzu2016} stays stable for $\alpha\leq0.6$.  In this solid, the molecules arrange into tubes and the hydrogen bonds tend to point towards the oxygens from the same tube (see Appendix~\ref{A3}). The hexatic phase observed for planar confinement at $\rho$ = 1.47 g~cm$^{-3}$ and high temperatures demands a deeper analysis to verify whether the orientational long-range order and translational short-range order are kept under the different values of the amplitude of the modulation. 

\subsubsection{Hexatic phase}

As previously mentioned, the triangular modulation with the lattice parameter $a$ = 4.78 $\text{\AA}$ is expected to disfavour highly triangular structured phases, and therefore, the hexatic phase. In addition, one expects that a modulation inserting a lattice of energy maxima and minima along the plane should hinder continuous shear, and thus constrain the appearance and diffusion of the dislocations, responsible for the lack of long-range translational order of the hexatic phase~\cite{kosterlitz1973,young1979,halperin1978,nelson1979,von2007}. These dislocation appear when the density of the ideal triangular ice (expected to be at $\rho$ = 1.76 g~cm$^{-3}$~\cite{zubeltzu2016}) is decreased.

For the characterization of the hexatic phase, we also calculate the oxygen-oxygen first-neighbor correlation function (C$_{\text{OO}}$), as was already used in~\cite{zubeltzu2016}. From C$_{\text{OO}}$, we calculate the characteristic time $\tau$ of the C$_{\text{OO}}$ curve, which is defined by C$_{\text{OO}}(\tau) =$ 0.5. For $\alpha\leq0.4$, the oxygens arrange themselves into a triangular lattice showing similar structural and dynamical features as the ones observed in the hexatic phase at $\alpha$ = 0~\cite{zubeltzu2016}: similar RDFs (see Appendix~\ref{A4}), diffusivities ($D =$ 6~10$^{-7}$~cm$^2$s$^{-1}$ at $T$ = 300 and $\alpha$ = 0.4 versus $D =$ 9.5~10$^{-7}$~cm$^2$s$^{-1}$ for $\alpha$ = 0.0~\cite{zubeltzu2016}) and oxygen-oxygen first-neighbor correlation functions [Fig.~\ref{fig:cor}~(a)]. We also observe the existence of shear motion along the main directions of the triangular lattice in the oxygen $xy$ density histogram [Fig.~\ref{fig:cor}~(b)]. All these indicators combined suggest that the hexatic phase remains stable for $\alpha\leq0.4$. At $\alpha$ = 0.6, however, although there is still a clear structuring of the oxygens into a triangular lattice, we do not observe shear motion of oxygens [Fig.~\ref{fig:cor}~(c)]. The increase in the diffusivity ($D$ = 3~10$^{-6}$~cm$^2$s$^{-1}$), decrease of the characteristic time of the C$_{\text{OO}}$ function in Fig.~\ref{fig:cor}~(a), and the less structured RDF (see Appendix~\ref{A4}) suggest that the hexatic phase is sufficiently frustrated to be no longer stable under these conditions. Although the triangular structuring is still present in the $xy$ density histogram of the oxygens, water seems to behave as a dense triangular liquid. Therefore, the indicators employed in this section suggest that the continuous melting reported in~\cite{zubeltzu2016} washes out for $\alpha\geq0.6$.


\subsubsection{Intercalated honeycomb ice}

The structural and dynamical properties of the honeycomb ice, square tubes ice, and the triangular phases appearing in Fig.~\ref{fig:PD} have been already well described in previous works~\cite{corsetti2015b,zubeltzu2016} and $\alpha$ does not affect them significantly (see Appendices~\ref{A2} and \ref{A5}). However, when $\alpha$ = 0.8 and 1.0 a different form of ice stabilizes at $\rho$=1.47 g~cm$^{-3}$, which we call intercalated honeycomb ice. Fig.~\ref{fig:ice} shows the $xy$ density histogram of oxygens and the $z$ density profile characteristic of this solid. The oxygen atoms arrange themselves into three layers: the two exterior layers conform a AA stacked honeycomb structure favoured by the energy landscape produced by the Lennard-Jones confining particles. The oxygens from the central layer fill the centre of the honeycomb hexagons coinciding with the $xy$ coordinates of the confining Lennard-Jones particles. Between the optimal densities of the honeycomb ice ($\rho$=1.17 g~cm$^{-3}$), and the intercalated honeycomb ice ($\rho$=1.47 g~cm$^{-3}$), we observe regions of coexistence of both phases independently of the temperature (see Appendix~\ref{A2}). This is a signature of having a first-order phase transition in a $NVT$ ensemble.

\begin{figure}[t!]
\centering
\subfigure[\ ]{\includegraphics[width=0.38\textwidth]{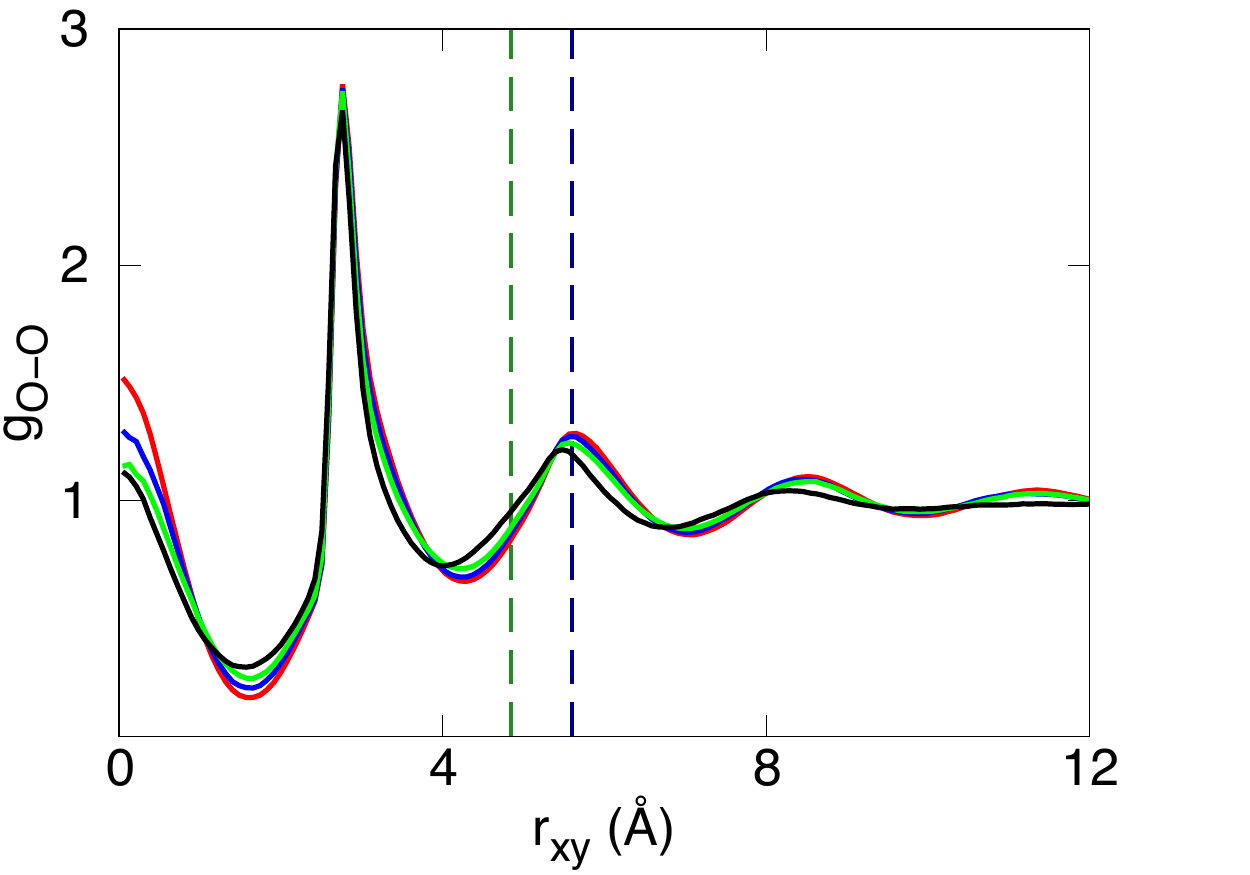}}\\
\subfigure[\ ]{\includegraphics[width=0.40\textwidth]{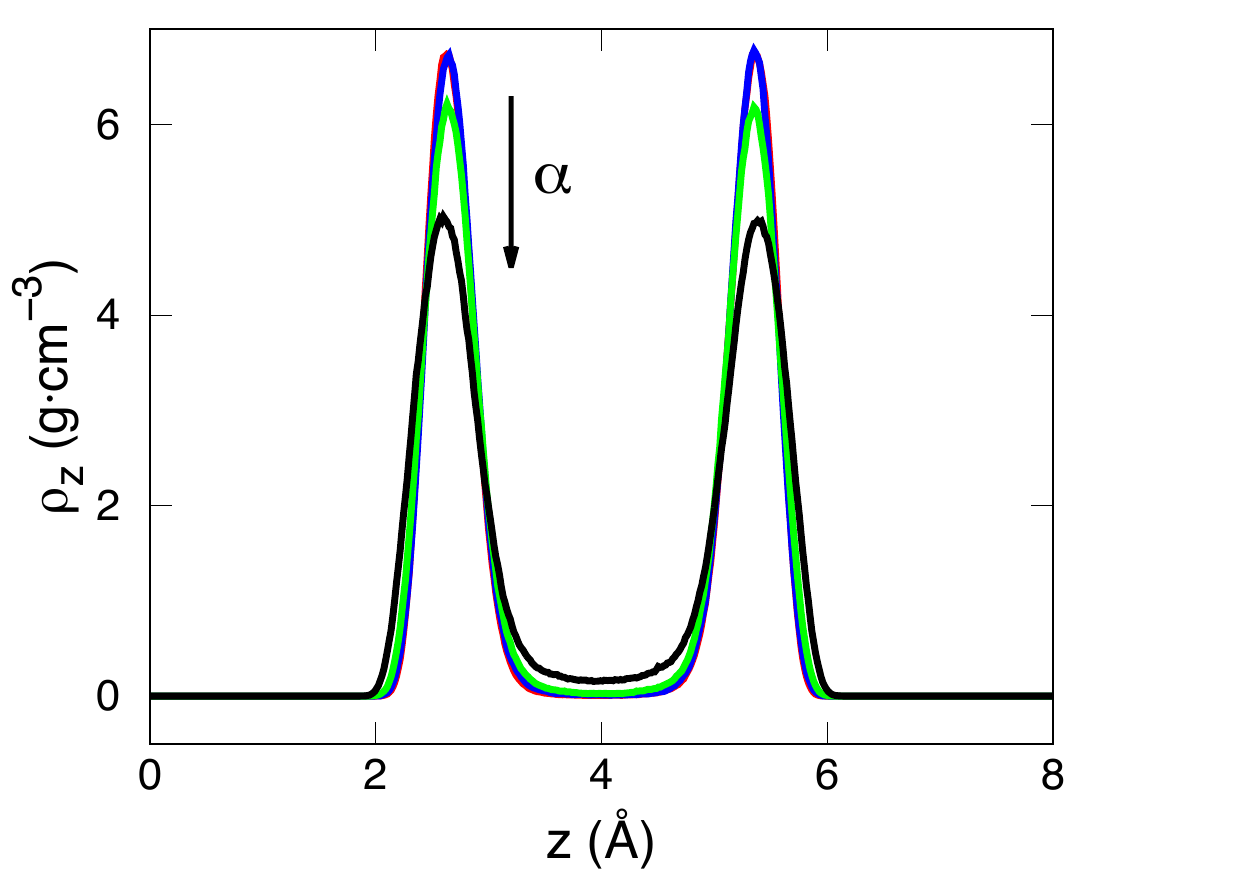}}
\caption{(a) Oxygen-oxygen radial distribution function, and (b) $z$ density profile both at $T$ = 300 K, and $\rho$ = 1.37 g~cm$^{-3}$ for $\alpha$ = 0 (red), 0.2 (blue), 0.4 (green), and 0.6 (black). The vertical dashed lines in (a) highlight the second (green) and third (blue) neighbor peaks in triangular ice RDF~\cite{zubeltzu2016}, the latter absent in honeycomb ice.}
\label{fig:liquid}
\end{figure}

\subsubsection{Liquid}

The effect of the corrugation on the structure of the liquid is different depending on its density. For densities ranging between $\rho$ = 0.97$-$1.17 g~cm$^{-3}$, the liquid rapidly freezes at the highest temperatures considered as $\alpha$ increases, as shown in Fig.~\ref{fig:PD}. Before it freezes, however, the liquid does not show appreciable changes on its structure (see Appendix~\ref{A6}). At higher densities, the liquid resists to freeze, specially at $\rho$ = 1.37 g~cm$^{-3}$, which only freezes after $\alpha$ = 0.8. Fig.~\ref{fig:liquid} shows the RDFs, and $z$ density profiles of liquid water at $T$ = 300 K, and $\rho$ = 1.37 g~cm$^{-3}$ for different values of $\alpha$. The vertical dashed lines on Fig.~\ref{fig:liquid}(a) highlight the position of the second and third neighbor peaks in triangular ice $r_h$ = $\sqrt{3}r_{\text{O-O}}$ = 4.78 $\text{\AA}$ and $r_t$ = 2$r_{\text{O-O}}$ = 5.52 $\text{\AA}$, respectively, the latter being absent in honeycomb ice~\cite{zubeltzu2016}. With no corrugation, the liquid at this density shows structural features on the RDF and density profile characteristic of triangular ice~\cite{zubeltzu2016}: high inter-planar correlation (high value on $g_{\text{O-O}}(0)$), a pronounced peak on the RDF at $r_t$, and two pronounced peaks on the density profile with little density and flux of molecules between them. When $\alpha$ is applied, the RDFs barely increase their value at $r_h$, which shows the clear resistance of the liquid to maintain the triangular structure despite the fact that it is disfavoured by the external modulation. The liquid, instead of becoming more honeycomb-like, shows a destructuring effect under the modulation (Fig.~\ref{fig:liquid}): as $\alpha$ increases, the RDFs show smaller peaks and decrease their value at the origin that measures the inter-layer correlation. The two peaks on the density profile decrease while the flux of molecules between them increases. The destructuring tendency with the external modulation is also supported by the observation of the disappearance of the solid-liquid coexistence point observed at $T$ = 220 K, $\rho$ = 1.37 g~cm$^{-3}$, and $\alpha$ = 0, and 0.2 in the phase diagram (Fig.~\ref{fig:PD}).

\begin{figure}[b]
\includegraphics[width=0.5\textwidth]{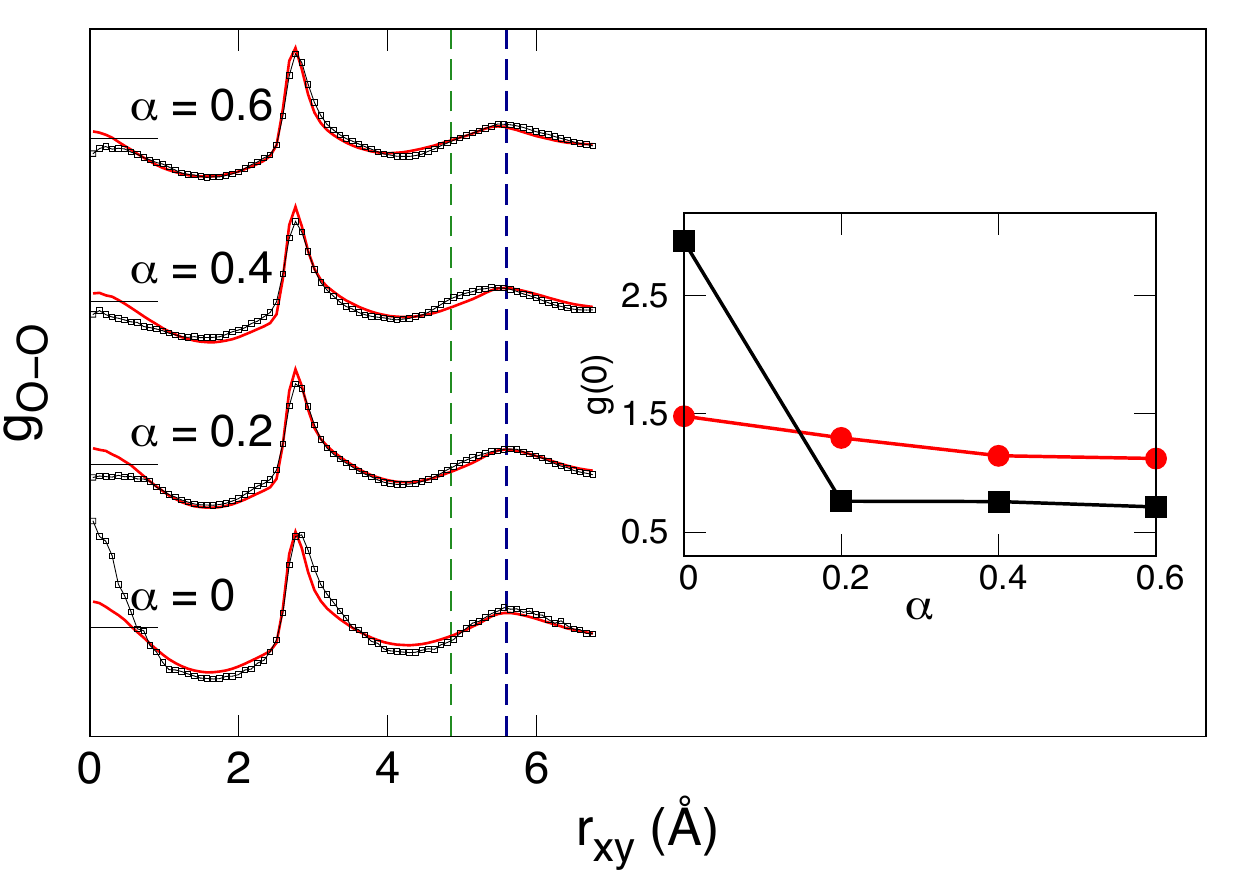}
\caption{Oxygen-oxygen radial distribution functions in the $xy$ plane at $T$ = 300 K, and $\rho$ = 1.37 g~cm$^{-3}$ for different values of $\alpha$. The curves are shifted on the $y$ axis; the value of saturation of each curve is marked by a horizontal line. The continuous lines correspond to the FFMD calculations while the line-points curves to the AIMD calculations. The vertical dashed lines highlight the second (green) and third (blue) neighbor peaks in triangular ice (dark blue) RDF~\cite{zubeltzu2016}, the latter absent in honeycomb ice. The inset shows the value at the origin of the RDFs obtained by FFMD calculations (red circles), and AIMD calculations (black squares).}
\label{fig:rdf1}
\end{figure}

We compare the effect of the corrugated wall on the liquid with FFMD calculations with the one obtained with AIMD calculations. The red points in Fig.~\ref{fig:PD} represent the density, temperature, and $\alpha$ values for which AIMD calculations have been done. The direct comparison of the RDFs obtained at $\rho$ = 1.37 g~cm$^{-3}$ and $T$ = 300 K (Fig.~\ref{fig:rdf1}) show the good agreement between the structural results obtained by the two different methods. Previous works~\cite{corsetti2015b,zubeltzu2016} with planar confinement, together with the ones shown here, indicate that the results obtained with the TIP4P/2005 empirical force-field and the {\em ab inito} molecular dynamics with the vdW-DF$^{\text{PBE}}$ functional agree surprisingly well.

The small differences between the RDFs obtained by AIMD and FFMD calculations agree with what presented in previous reports: the RDFs obtained by AIMD are more structured, showing larger correlation peaks positioned on the O-O distances characteristic of the triangular ice. Both calculation methods show that the liquid is triangularly structured independently of the density. It is particular of the AIMD RDFs however, that the activation of the modulation affects significantly the maximum at $r_{\text{xy}}$ = 0, which measures the interplanar on-top correlation. The peak at the origin of the RDF is characteristically pronounced for a triangular liquid where there is a tendency for AA stacking~\cite{zubeltzu2016}. The peak decreases abruptly for AIMD when $\alpha$ = 0.2 for all the sampled densities. In order to understand this behavior, we calculate the interlayer RDF, which takes into account the correlation of an oxygen atom with the oxygens of the other layer. 

\begin{figure}[t!]
\centering
\subfigure[\ ]{\includegraphics[width=0.38\textwidth]{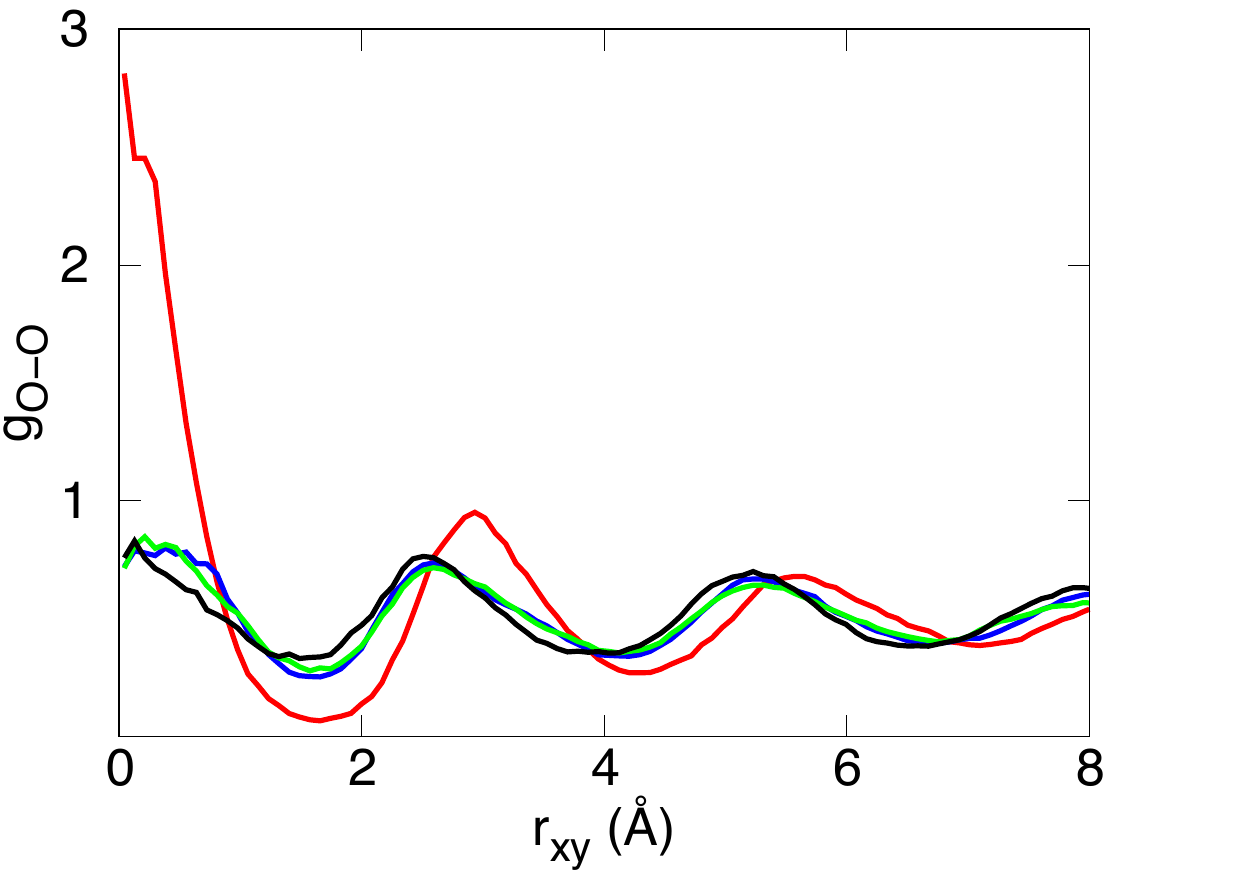}}\\
\subfigure[\ ]{\includegraphics[width=0.40\textwidth]{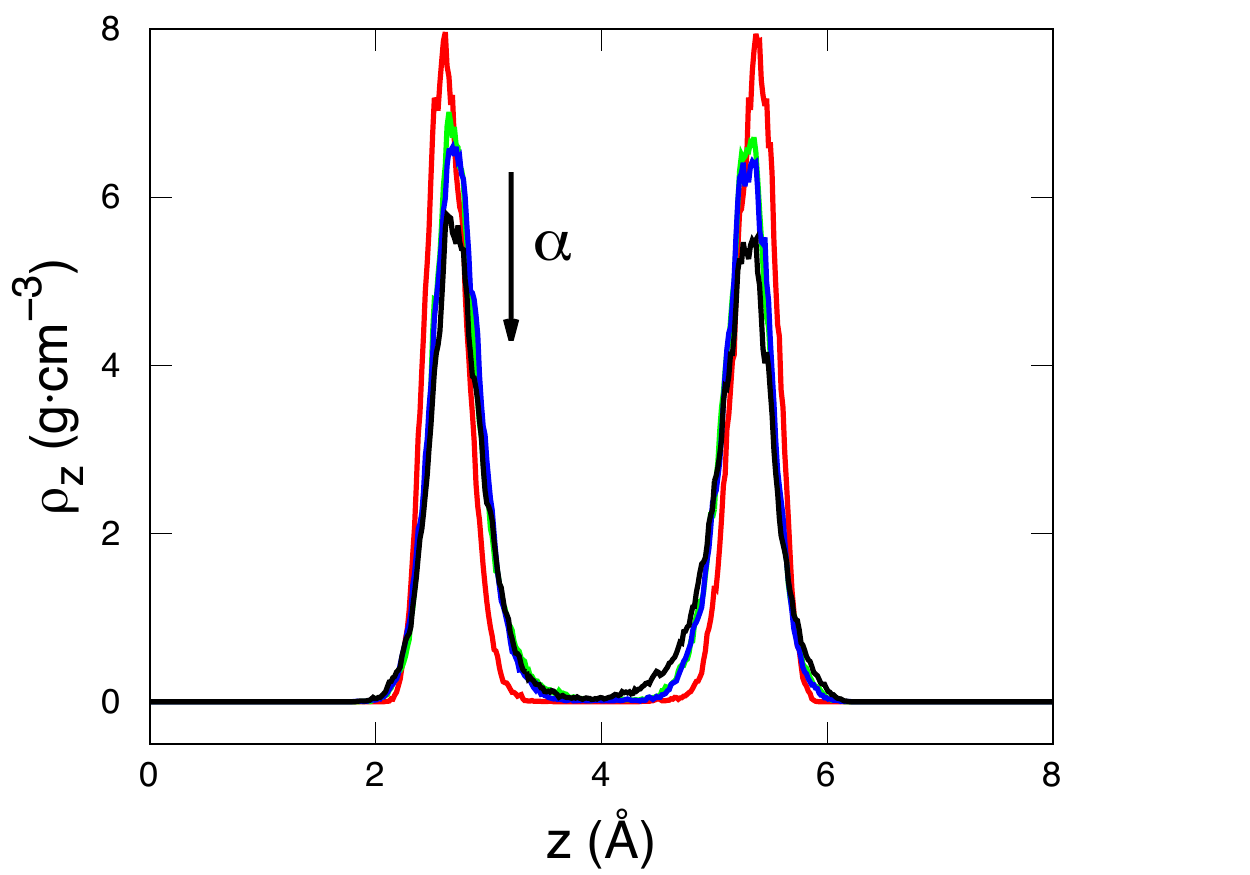}}
\caption{(a) Oxygen-oxygen interlayer radial distribution function, and (b) $z$ density profile both at $T$ = 300 K and $\rho$ = 1.37 g~cm$^{-3}$ for $\alpha$ = 0 (red), 0.2 (blue), 0.4 (green), and 0.6 (black).}
\label{fig:AIMD}
\end{figure}

Fig.~\ref{fig:AIMD} shows the interlayer RDFs and density profiles at $T$ = 300 K and $\rho$ = 1.37 g~cm$^{-3}$ for different modulation amplitudes obtained by AIMD calculations. The $\rho(z)$ density profiles show that once the corrugation is activated the two water layers get noticeably closer to each other. The intensity of the peaks decreases as the amplitude of the modulation increases. As the AIMD liquid is more triangularly structured than the FFMD one, and the external modulation frustrates triangularly structured phases, the effect of the modulation in the AIMD calculation is greater than in the FFMD case. Therefore,  the oxygens closer to maxima of the modulation tend to locate closer to the center of the film displacing the peaks of the density profiles. This displacement explains the loss of AA stacking: the distance between the oxygen layers becomes smaller than the optimal distance for a vertical H-bond, and therefore, the vertical bonds get tilted with respect to the $z$ axis. The effect on the stacking is also seen by the slight differences of the peak away from $r_{\text{xy}}=0$ in Fig~\ref{fig:AIMD}~(a), which is accompanied by a small displacement of the other peaks towards smaller distances, reflect of the change in stacking correlations.

The obtained diffusivities for the liquid and solid phases for all the values of $\alpha$ are very similar to the ones previously calculated in the absence of modulation~\cite{zubeltzu2016} and observed for bulk water~\cite{corsetti2013}: At $T$ = 300 K, the liquid has a diffusivity of the order of $D$ $\sim$ $10^{-5}$ cm$^2$s$^{-1}$ while the solid phases $D$ $\sim$ $10^{-8}-10^{-9}$ cm$^2$s$^{-1}$. The calculations of the mean-square displacement (MSD) at the different points of the phase diagrams show that for a given $T$ and $\rho$, the diffusivity does not significantly change with $\alpha$, as long as a phase transition does not occur (see Appendix~\ref{A6}).

\subsection{Realistic lattice parameters}

After analyzing a lattice parameter that is ideally commensurated with honeycomb ice, we study the effect of corrugated walls with more realistic lattice parameters. We chose three different triangular lattices with $a$ = 2.50, 2.75, and 3.00 $\text{\AA}$ that cover a range of typical values for closed-packed metal surfaces, such as Ni (2.49 $\text{\AA}$), Cu (2.56 $\text{\AA}$), Pt (2.78 $\text{\AA}$), Au (2.88 $\text{\AA}$), and Ag (2.89 $\text{\AA}$). In this section we restrict our simulation to $T$ = 300 K, which corresponds to the highest $T$ in Fig~\ref{fig:PD}. 

\subsubsection{Structure}

\begin{figure}[b]
\includegraphics[width=0.5\textwidth]{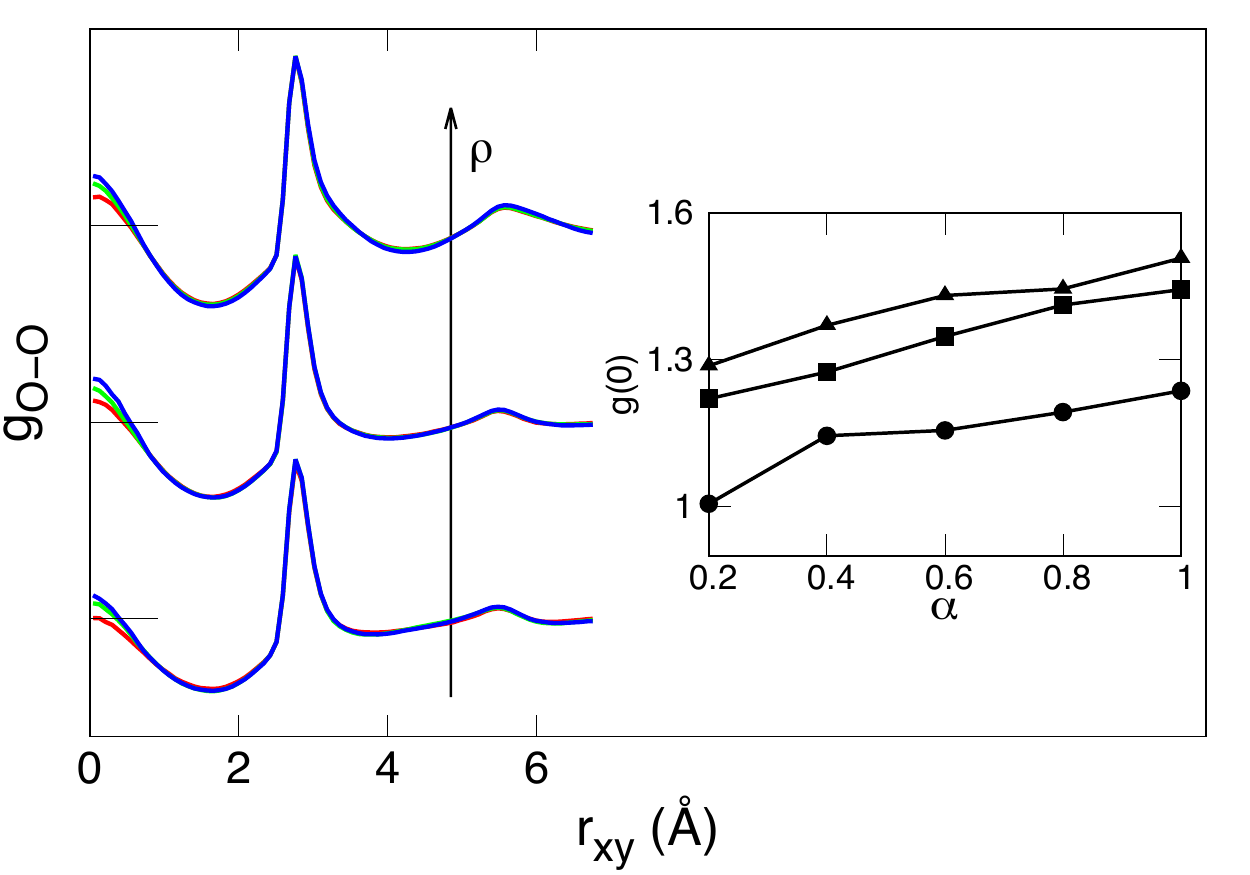}
\caption{RDFs for $a$ = 3.0 $\text{\AA}$ at $\rho$ = 1.17 (below), 1.27 (middle), and 1.37 (above) g~cm$^{-3}$ at $\alpha$ = 0.2, 0.6, and 1. The curves are shifted on the $y$ axis; the value of saturation of each curve is marked by a horizontal finite line. The inset shows how $g_{\text{O-O}}(0)$ changes for each value of density with respect to $\alpha$.}
\label{fig:rdf2}
\end{figure}

We carry out FFMD calculations for the same densities as in the previous section, and $\alpha$ = 0.2, 0.4, 0.6, 0.8, and 1.0. We observe no significant change in the different phases under the effect of the corrugation, staying liquid at low densities and triangularly structured at high densities independently of $\alpha$ [as in Fig.~\ref{fig:PD}(a) at $T$ = 300 K]. The high-density triangular phase shows the features of the hexatic phase previously mentioned: similar RDFs, and a clear triangular lattice in the $xy$ density histograms of the oxygens with the usual shear motions along the main directions of the lattice (see Appendix~\ref{A7}). The liquid shows almost no change in the RDFs and in the density profiles, when varying $\alpha$. Fig.~\ref{fig:rdf2} shows the RDFs for $a$ = 3.0 $\text{\AA}$ at $\rho$ = 1.17, 1.27, and 1.37 g~cm$^{-3}$ (different vertical shifts for each $\rho$), and $\alpha$ = 0.2, 0.6, and 1.0. When $r_{\text{xy}} >$ 1.0 $\text{\AA}$, the RDFs with the same density and different values of $\alpha$ are almost indistinguishable from each other. The peak around $r_{\text{xy}}$ = 0 $\text{\AA}$, which contains information about the interlayer on-top correlation, shows a small tendency to increase with $\alpha$ (inset in Fig.~\ref{fig:rdf2}) the opposite to the behavior for the ideal $a$. This means that as the corrugation increases, the liquid tends to structure into better AA stacking. From these results we deduce that for $a$ = 3.0 $\text{\AA}$, although the relative distances among oxygen atoms are almost unaffected by the corrugation, the water molecules from one layer tend to be on top of another from the other layer as the corrugation increases. The figures are very similar for $a$ = 2.5, and 2.75 $\text{\AA}$ and the same conclusions can be drawn for them (see Appendix~\ref{A8}).

\begin{figure}[t]
\includegraphics[width=0.5\textwidth]{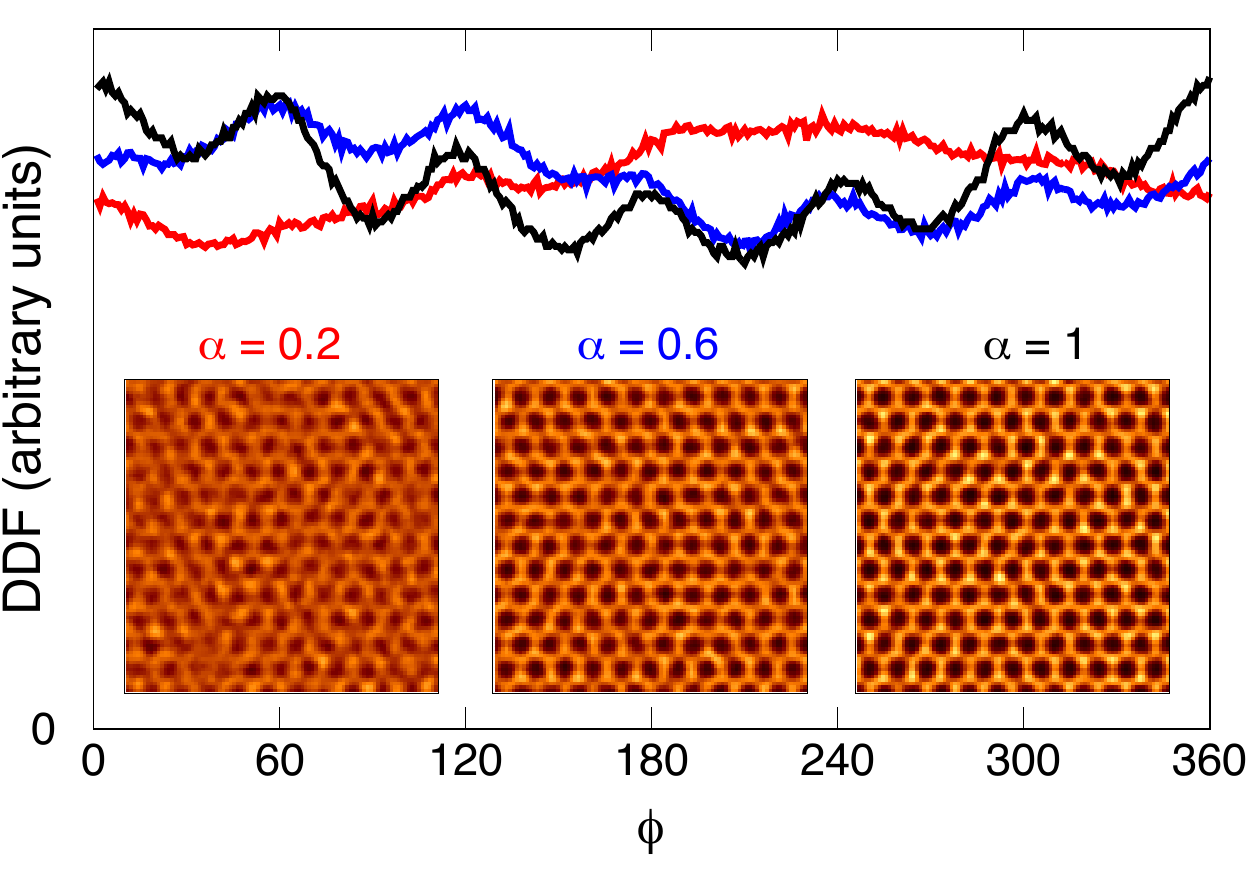}
\caption{Dipolar angle distribution function of the water molecules for $a$ = 3.0 $\text{\AA}$ at $T$ = 300 K, $\rho$ = 1.27 g~cm$^{-3}$, and $\alpha$ = 0.2 (red), 0.6 (blue), and 1.0 (black). The insets show the $xy$ density histogram of the oxygens at the three values of $\alpha$: the brighter the color, the larger the density (see text for color scheme).}
\label{fig:ddf}
\end{figure}

Although the RDFs and density profiles of oxygens suggest that the structure of the liquid is barely affected by the corrugation, the $xy$ density histogram of the oxygens shows clear effects on the structure of the liquid. The insets in Fig.~\ref{fig:ddf} show the oxygen $xy$ density histogram for $a$ = 3.0 $\text{\AA}$, at $\rho$ = 1.27 g~cm$^{-3}$, and $\alpha$ = 0.2, 0.6, 1.0. When the corrugation is activated, the oxygens avoid being close to the Lennard-Jones confining particles due to the effective repulsive force they represent, and they are structured anisotropically, resulting in the heterogeneous images shown in the insets of Fig.~\ref{fig:ddf}. The maximum value of the density oscillations in the insets are $138\%$, $186\%$, and $237\%$ of the mean value of the density for $\alpha$ = 0.2, 0.6, and 1.0, respectively. To quantify the effect of this structuring of the liquid, we calculate the dipole distribution function (DDF): we average the distribution of the polar angle (projected in the $xy$ plane) of the molecular dipoles. Fig.~\ref{fig:ddf} shows the DDFs for $a$ = 3.0 $\text{\AA}$, at $\rho$ = 1.27 g~cm$^{-3}$, and $\alpha$ = 0.2, 0.6, 1.0. We can clearly observe that as $\alpha$ increases, the DDFs show six pronounced peaks in the multiples of 60$^{\circ}$. This anisotropic structuring effect is observed for all sampled points of the liquid phase, and it is greater as $\rho$, and $a$ are increased. These results show that the (not surprising) anisotropy in the molecular density of the liquid has a surprisingly small effect in the liquid structure (correlation) and dynamics (diffusivity), in spite of the orientational effect.

\begin{figure}[b]
\includegraphics[width=0.48\textwidth]{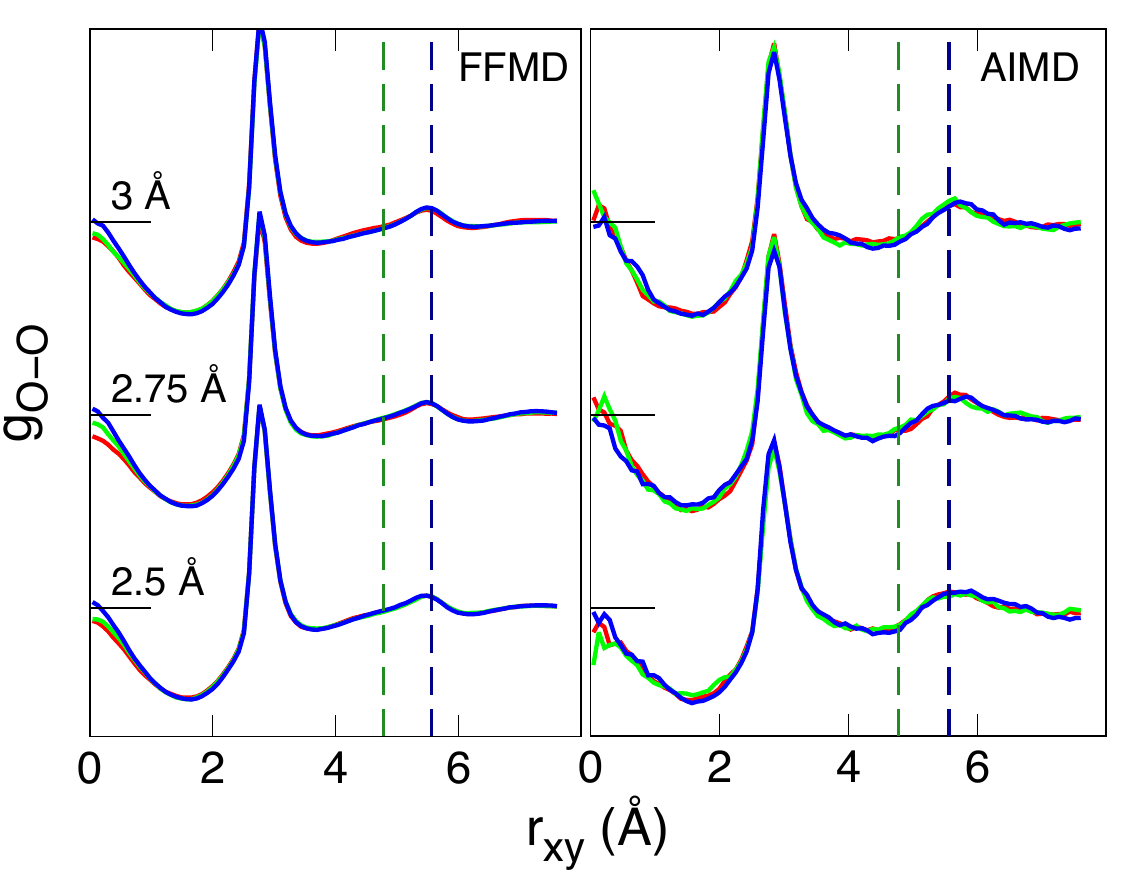}
\caption{RDFs obtained by FFMD (left), and AIMD (right) calculations at $T$ = 300 K, and $P_{\parallel}$ = 0. A different height is given to the curves for each lattice parameter: $a$ = 2.5 (bottom), 2.75 (middle), and 3.0 $\text{\AA}$ (top). For each lattice parameter, there are three curves, each one corresponding to a different value of $\alpha$ = 0.2 (red), 0.6 (green), and 1.0 (blue). The value of saturation of each curve is marked by a horizontal line. The vertical dashed lines highlight the second (green) and third (blue) neighbor peaks in triangular ice (dark blue) RDF~\cite{zubeltzu2016}, the latter absent in honeycomb ice.}
\label{fig:rdf3}
\end{figure} 

We now compare the results obtained by FFMD and AIMD calculations at $T$ = 300 K, and $P_{\parallel}$ = 0, for different values of $\alpha$ and $a$. Fig.~\ref{fig:rdf3} shows the RDFs obtained by FFMD and AIMD calculations at $a$ = 2.50, 2.75, and 3.00 $\text{\AA}$, and $\alpha$ = 0.2, 0.6, and 1.0. The effect of the corrugation in the RDFs is as negligible for AIMD as it was for FFMD. From the direct comparison of the FFMD and AIMD RDFs we obtain the same conclusions to the ones in the previous section: both methods give similar structural features of the liquid, but with the difference that the AIMD RDFs tend to be more triangularly structured showing a larger peak at $r_{\text{t}}$ = 2$r_{\text{O-O}}$. This is also supported by the comparison of the density profiles $\rho(z)$ obtained by both calculation methods: although they are very similar, the ones obtained by AIMD show more pronounced peaks. These results agree with the results obtained for $a$ = 4.78 $\text{\AA}$ and previous works where the same empirical force-field and DFT functional~\cite{corsetti2015b,zubeltzu2016} were used. The differences between both methods are minor.

\subsubsection{Diffusivity}

As previously mentioned, with no corrugation ($\alpha$ = 0) the sampled liquid points of the phase diagram show diffusivities of the order of $D\sim10^{-5}$ cm$^2$s$^{-1}$, while for the hexatic phase $D\sim10^{-7}$ cm$^2$s$^{-1}$~\cite{zubeltzu2016}. When the corrugation is applied with these three lattice parameters, and different values of $\alpha$, the diffusivities barely change in the liquid and in the triangular phase (see Appendix~\ref{A8}). The small changes with $\alpha$ in the mean square displacements curves are within the noise of the signal and do not display any clear trend; therefore, we conclude that within our accuracy, the diffusivities are barely affected by the corrugation formed with these three lattice parameters.  

\section{Conclusions}

In this work we simulate two-dimensionally confined water between two corrugated walls using empirical potentials (TIP4P/2005) and first principles. We propose a periodic confining potential based on Lennard-Jones particles that play the role of a confining substrate. We control the lattice type, lattice parameter and amplitude of the corrugation.

For $a$ = 4.78 $\text{\AA}$, ideal for the stabilization of two layers of honeycomb ice, at low-mid densities, the honeycomb ice rapidly stabilizes within the phase diagram as the modulation amplitude increases. However, before the liquid freezes, its triangular structure, although disfavoured by the external modulation, remains almost unchanged, independently of its density. This is supported by AIMD results.  At high densities and low temperatures, the square-tubes ice keeps being stable for $\alpha\leq$ 0.6. The hexatic phase appearing for planar confinement is destroyed for $\alpha\geq$ 0.6, which seems to liquify. When $\alpha\geq$ 0.8, a new solid phase appears, intercalated honeycomb ice. This solid is composed of two external honeycomb layers and a third middle layer with the molecules located in the centre of the honeycomb hexagons. The AIMD liquid tends to be slightly closer to triangular structure as previously reported for $\alpha$ = 0.0. The imposed modulation disfavours triangular-like structures, and therefore affects more significantly the AIMD liquid, which loses the AA stacking and well-defined layering shown in the planar confinement. 

For more realistic lattice parameters ($a$ = 2.50, 2.75, and 3.00 $\text{\AA}$) we do not observe significant changes in the phase behavior, staying liquid at low-mid densities, and triangular with similar features to the hexatic phase at high densities. Although the $xy$ density histograms of the oxygens and DDFs show that the molecules from the liquid display a significant anisotropy, there is hardly noticeable change in the structural (RDF) and dynamical (diffusivity) behavior. Finally, we have found that the differences between AIMD and TIP4P/2005 which we set out to explore, have shown to be significantly smaller than anticipated: after all, these structures are quite removed from the bulk liquid the forces were fitted to.

\section{Acknowledgements}
This work was partly funded by grants FIS2012-37549-C05 and FIS2015-64886-C5-1-P of the Spanish Ministerio de Economía, Industria y Competitividad, and Exp.\ 97/14 (Wet Nanoscopy) from the Programa Red Guipuzcoana de Ciencia, Tecnolog\'{i}a e Innovaci\'{o}n, Diputaci\'{o}n Foral de Gipuzkoa. We thank Jos\'{e} M. Soler and Fabiano Corsetti for useful discussions. The calculations were performed on the Arina HPC cluster (Universidad del Pa\'{i}s Vasco/Euskal Herriko Unibertsitatea, Spain) and MareNostrum (Barcelona Supercomputing Center). SGIker (UPV/EHU, MICINN, GV/EJ, ERDF and ESF) support is gratefully acknowledged.

\begin{appendices}
\section{Position of Lennard-Jones particles}
\label{A1}

We calculate the distance between the origin (divergence) of the confining wall and the Lennard-Jones confining particles $l_{z_1}$ by making the position of the minimum in $z$ of the mean total confining potential $\mean{U(x,y,z)}_{xy}$ independent from $\alpha$. Taking into account that a layer made of Lennard-Jones particles located at $z$ = $l_{z_1}$ integrates into the Lennard-Jones 10-4 potential, $\mean{U_{12\text{-}6}(x,y,z-l_{z_1})}_{xy}=U_{10\text{-}4}(z-l_{z_1})$, the total mean confining potential is:

\begin{figure}[t]
\centering
\subfigure[\ ]{\includegraphics[width=0.23\textwidth]{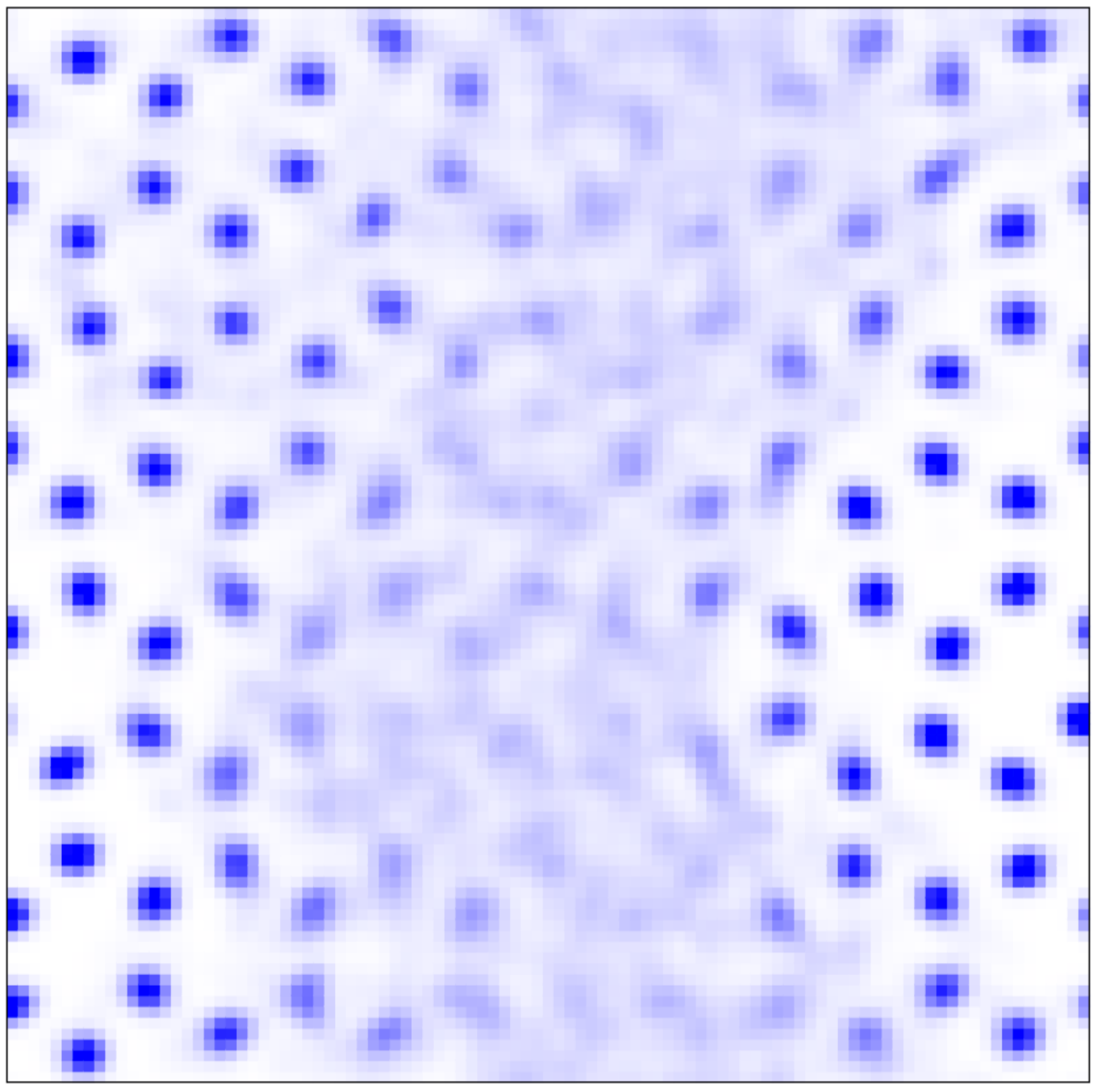}}
\subfigure[\ ]{\includegraphics[width=0.23\textwidth]{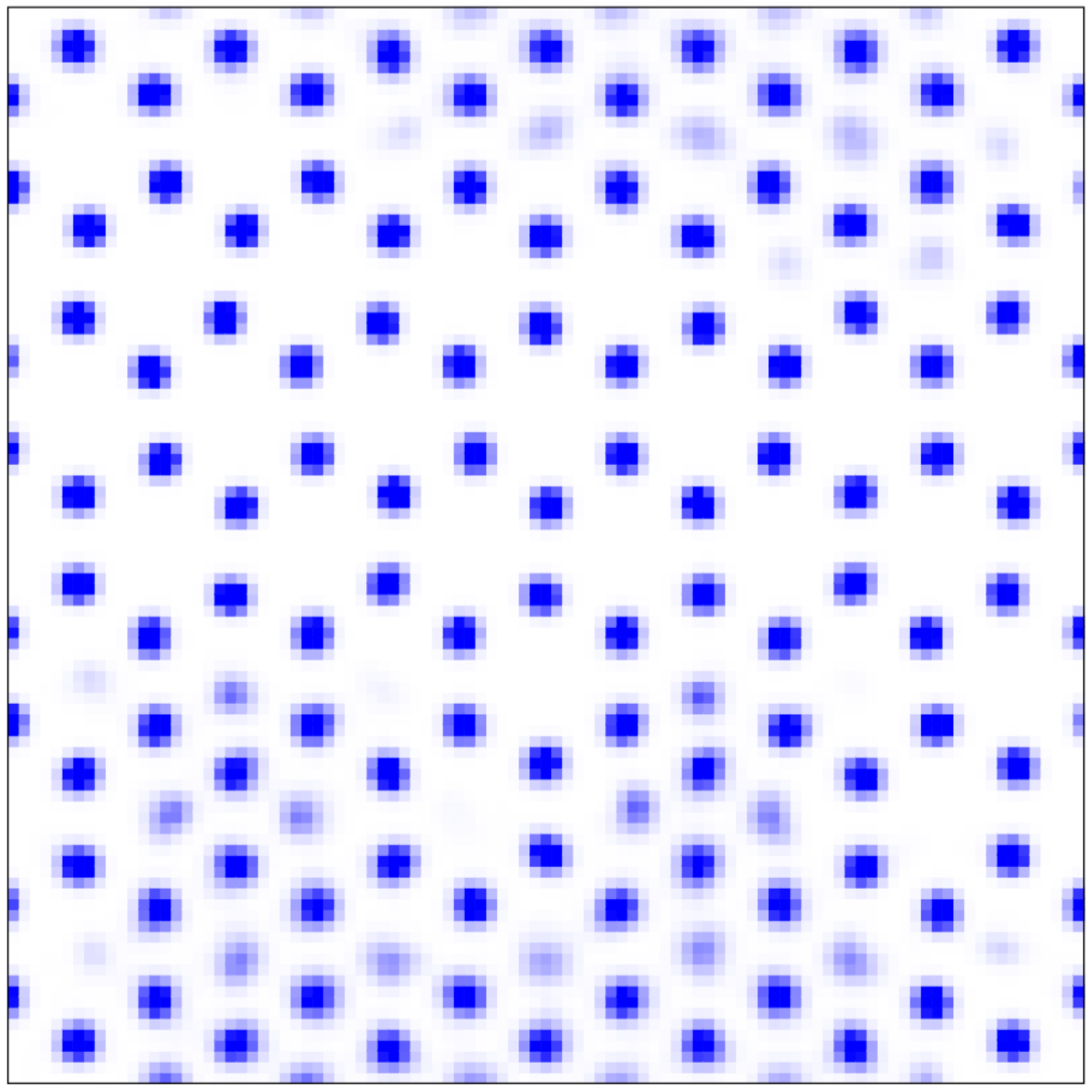}}\\
\caption{$xy$ density histogram of oxygen (blue) atoms averaged over 1 ns at (a) $T$ = 260 K, $\rho$ = 1.27 g~cm$^{-3}$, and $\alpha$ = 0.4, and (b) $T$ = 240 K, $\rho$ = 1.37 g~cm$^{-3}$, and $\alpha$ = 1.0.}
\label{fig:coex}
\end{figure}

\begin{equation}
\begin{split}
\mean{U(x,y,z)}_{xy}&=(1-\alpha)U_{9\text{-}3}(z)\\
&+\alpha\left[U_{10\text{-}4}(z-l_{z_1})+U_{9\text{-}3}(z-l_{z_2})\right].
\end{split}
\end{equation}
We then find the coordinate $z$ = $z_0$ at which the first term on the right has its minimum,
\begin{equation}
\dfrac{dU_{9\text{-}3}(z)}{dz}\Big|_{z_0}=0,
\end{equation}
and we obtain $l_{z_1}$ by imposing the second term on the right to have its minimum at the same coordinate $z_0$:
\begin{equation}
\dfrac{d\left[U_{10\text{-}4}(z-l_{z_1})+U_{9\text{-}3}(z-l_{z_2})\right]}{dz}\Big|_{z_0}=0.
\end{equation}
Following this procedure, the minimum of the total mean potential is located at $z_0$ and is independent from $\alpha$. 

\section{Phase coexistence for $a$ = 4.78 $\text{\AA}$}
\label{A2}

\begin{figure}[t]
\centering
\subfigure[\ ]{\includegraphics[width=0.23\textwidth]{tubes0}}
\subfigure[\ ]{\includegraphics[width=0.23\textwidth]{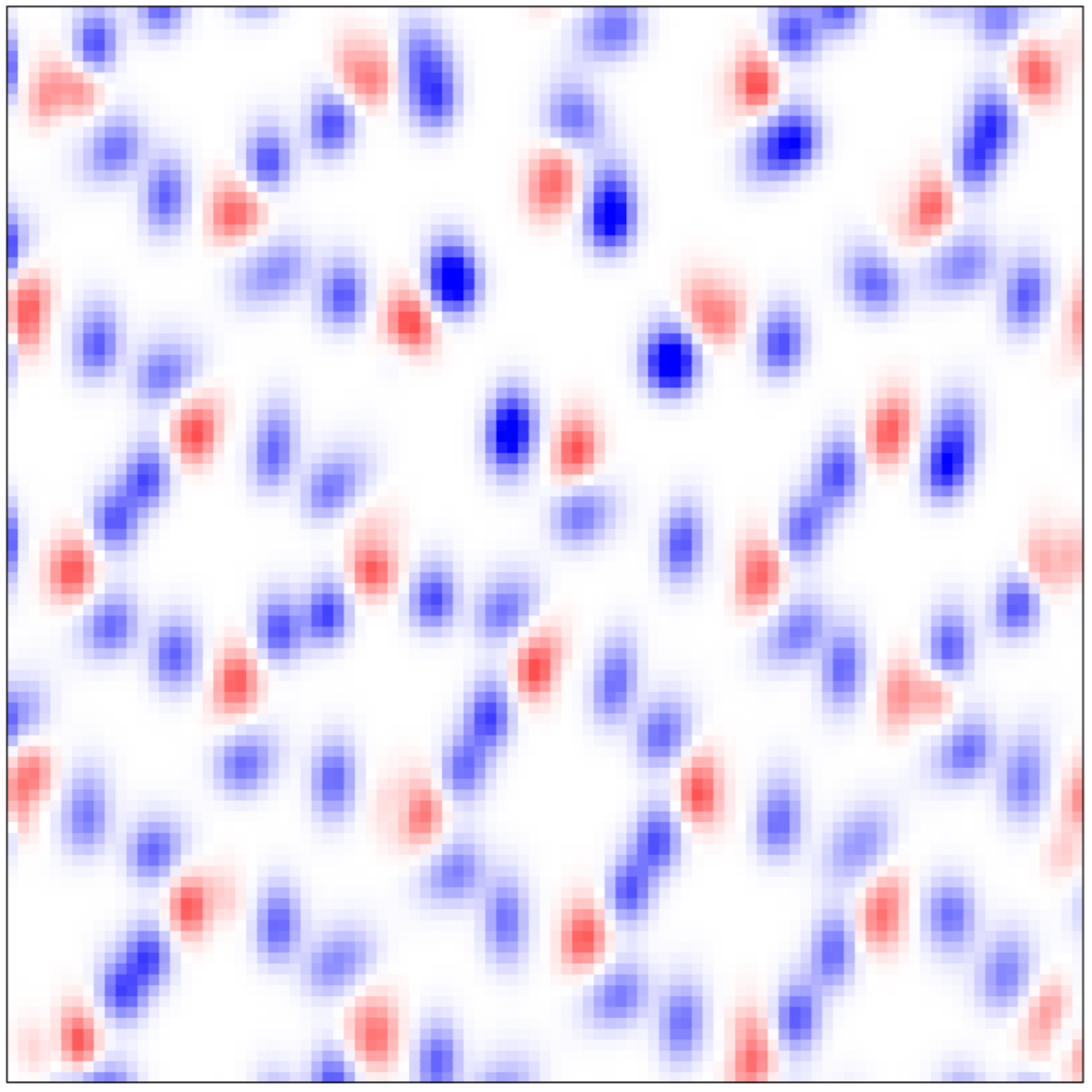}}\\
\caption{$xy$ density histogram of oxygen (red) and hydrogen (blue) atoms averaged over 1 ns at $T$ = 220 K, $\rho$ = 1.47 g~cm$^{-3}$, and $\alpha$ = 0 (a), 0.6 (b). All the images were obtained within a window of dimensions $\left ( 15 \times 15 \right)$~\AA$^2$ in the simulation cell.}
\label{fig:histos}
\end{figure}

\begin{figure}[t!]
\centering
\includegraphics[width=0.48\textwidth]{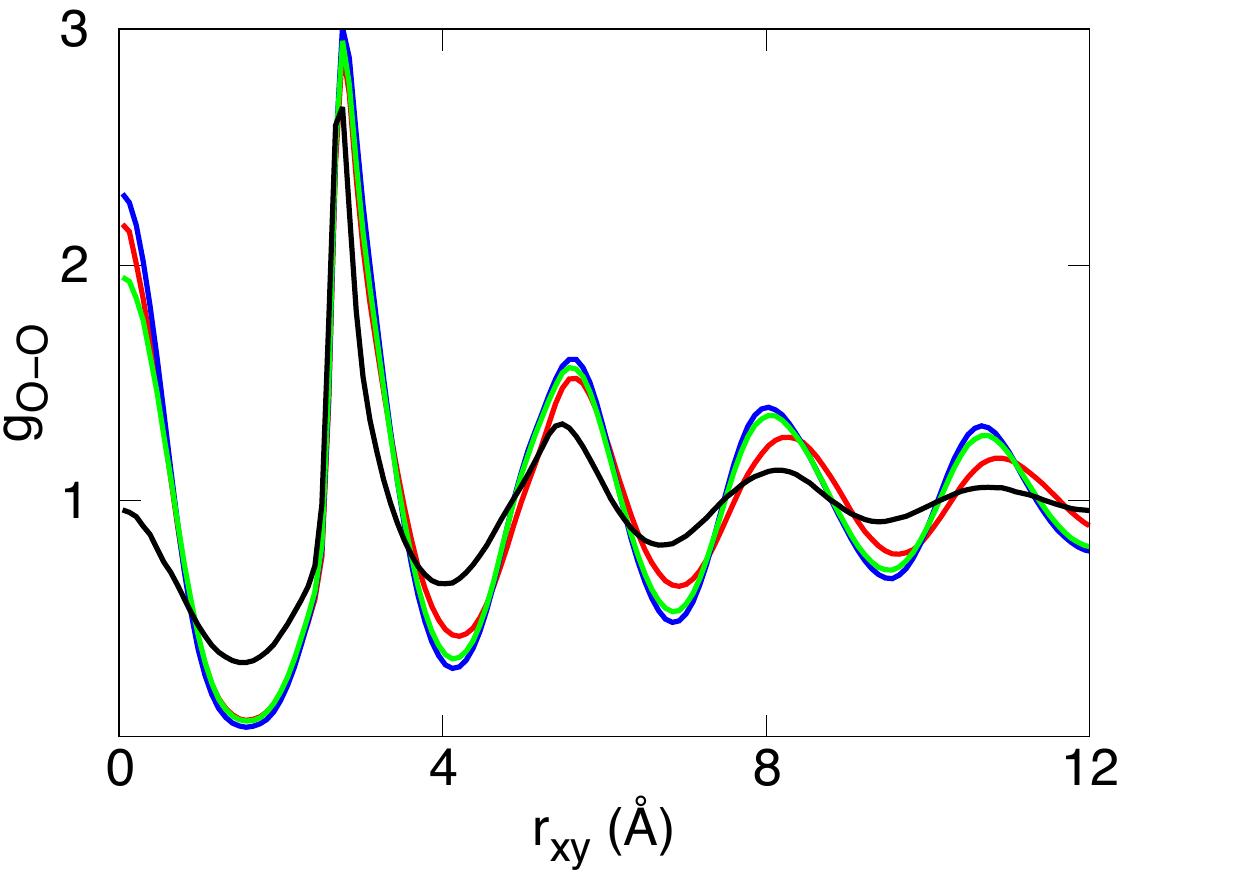}
\caption{Oxygen-oxygen radial distribution function at $T$ = 300 K and $\rho$ = 1.47 g~cm$^{-3}$ for $a$ = 4.78 $\text{\AA}$ and $\alpha$ = 0.0 (red), 0.2 (blue), 0.4 (green), and 0.6 (black).}
\label{fig:rdfhex}
\end{figure}

For $a$ = 4.78 $\text{\AA}$ we observe phase coexistence in some regions of the phase diagrams in Fig.~\ref{fig:PD}. Fig~\ref{fig:coex} shows the $xy$ density histogram of the oxygen atoms where the coexistence of the liquid-honeycomb and honeycomb-intercalated honeycomb phases are observed.

\begin{figure}[t]
\centering
\subfigure[\ ]{\includegraphics[width=0.23\textwidth]{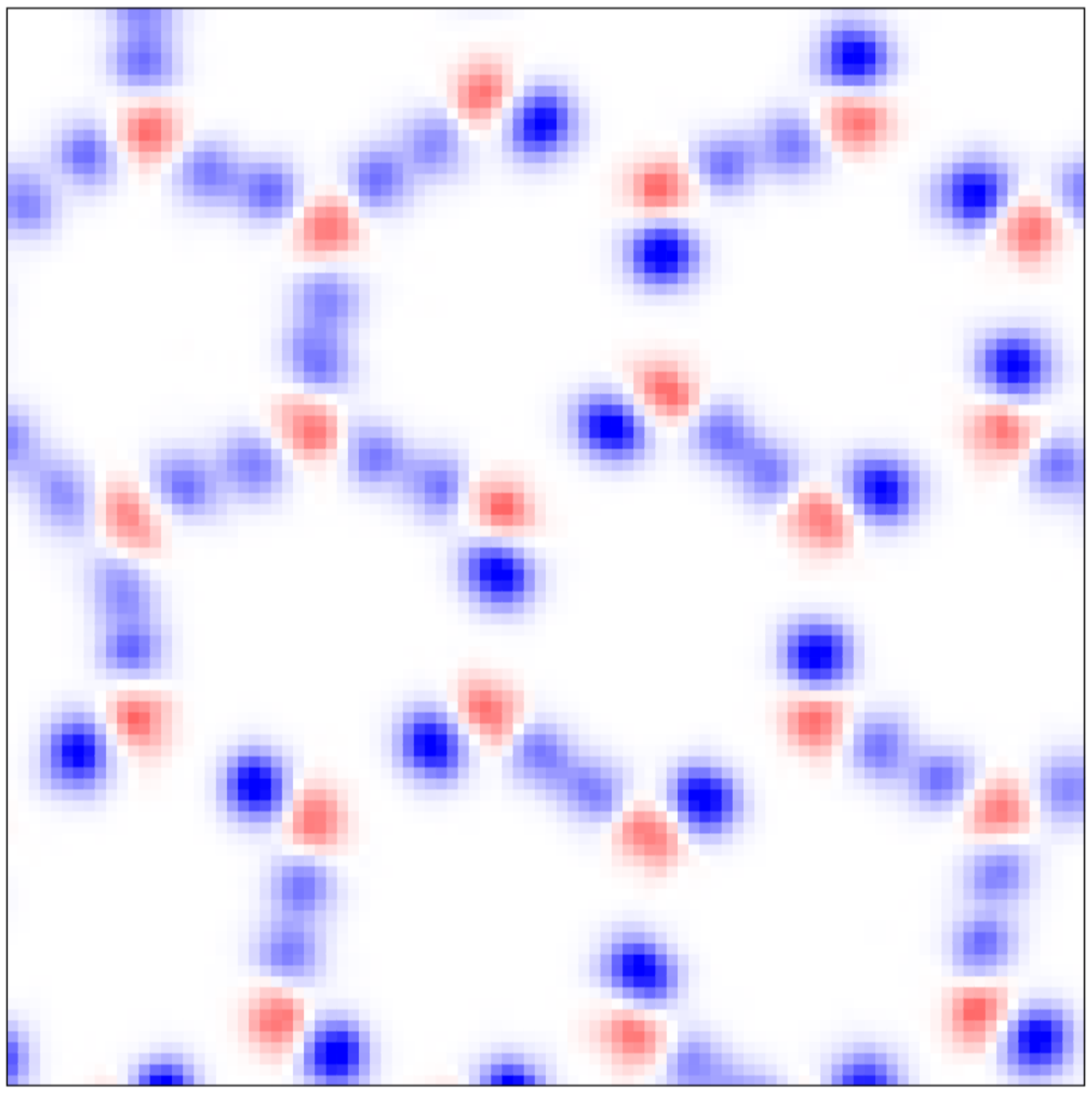}}
\subfigure[\ ]{\includegraphics[width=0.23\textwidth]{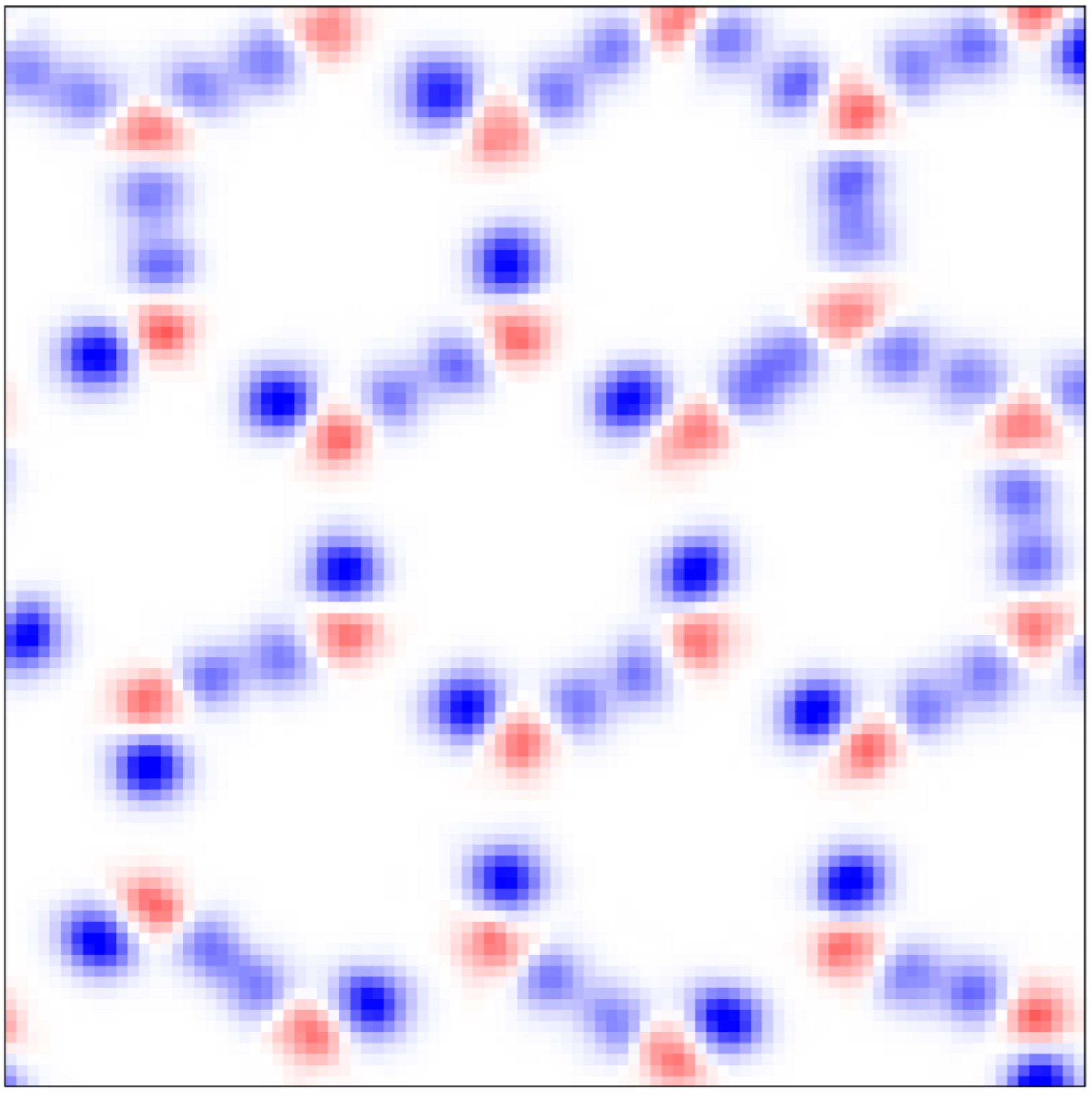}}\\
\caption{$xy$ density histogram of oxygen (red) and hydrogen (blue) atoms averaged over 1 ns at $\rho$ = 1.17 g~cm$^{-3}$, $T$ = 240 K, (a) $\alpha$ = 0.0 and (b) and $\alpha$ = 1.0. All the images were obtained within a window of dimensions $\left ( 15 \times 15 \right)$~\AA$^2$ in the simulation cell.}
\label{fig:histices}
\end{figure}

\begin{figure}[t!]
\centering
\subfigure[\ ]{\includegraphics[width=0.23\textwidth]{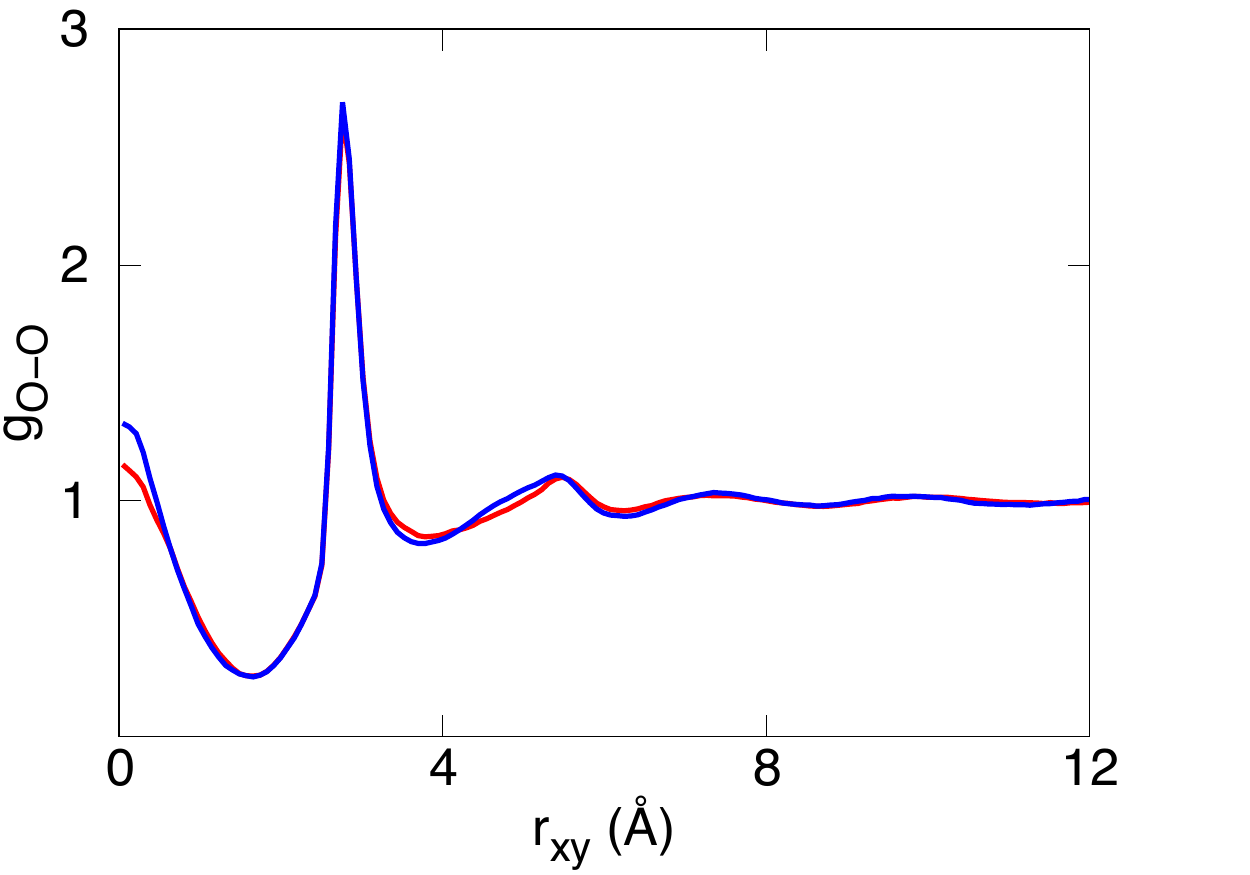}}
\subfigure[\ ]{\includegraphics[width=0.23\textwidth]{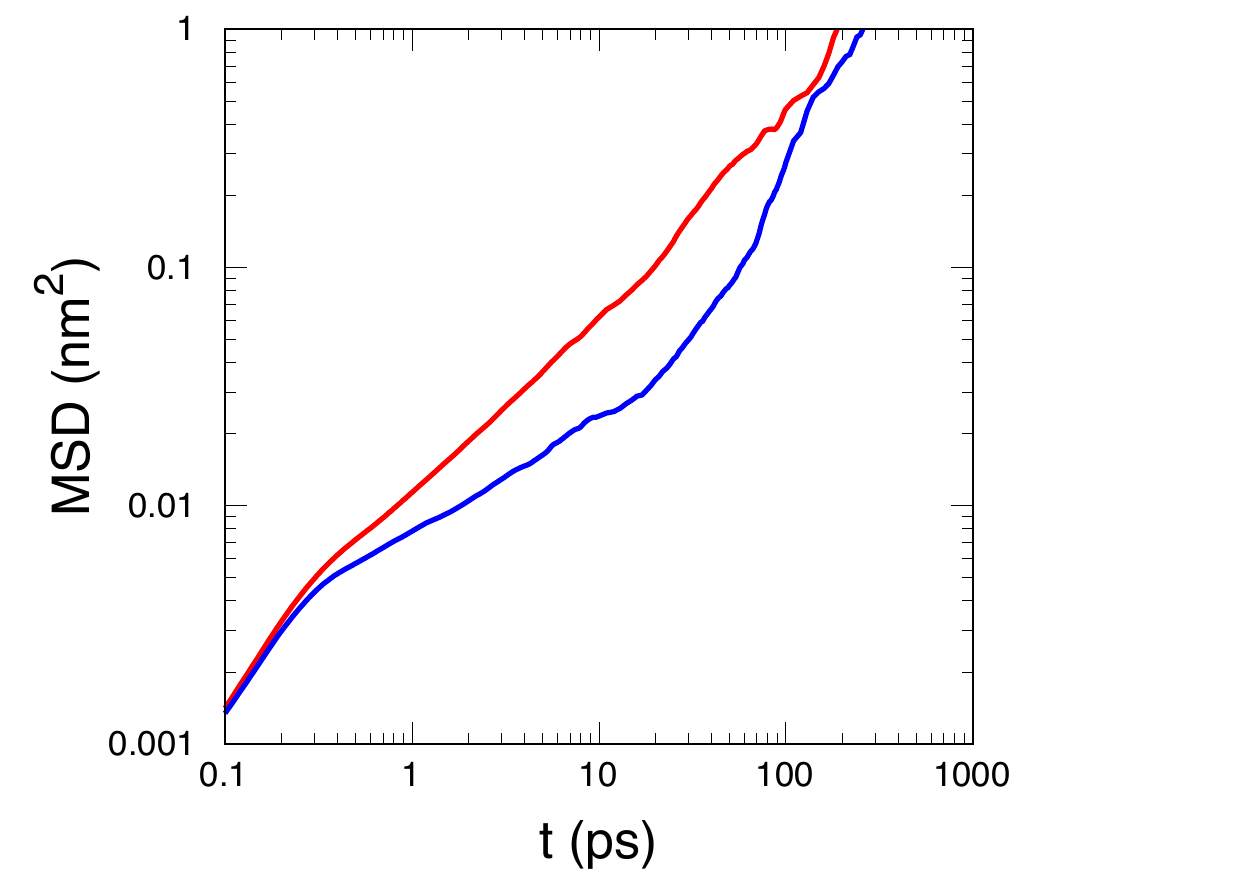}}\\
\subfigure[\ ]{\includegraphics[width=0.23\textwidth]{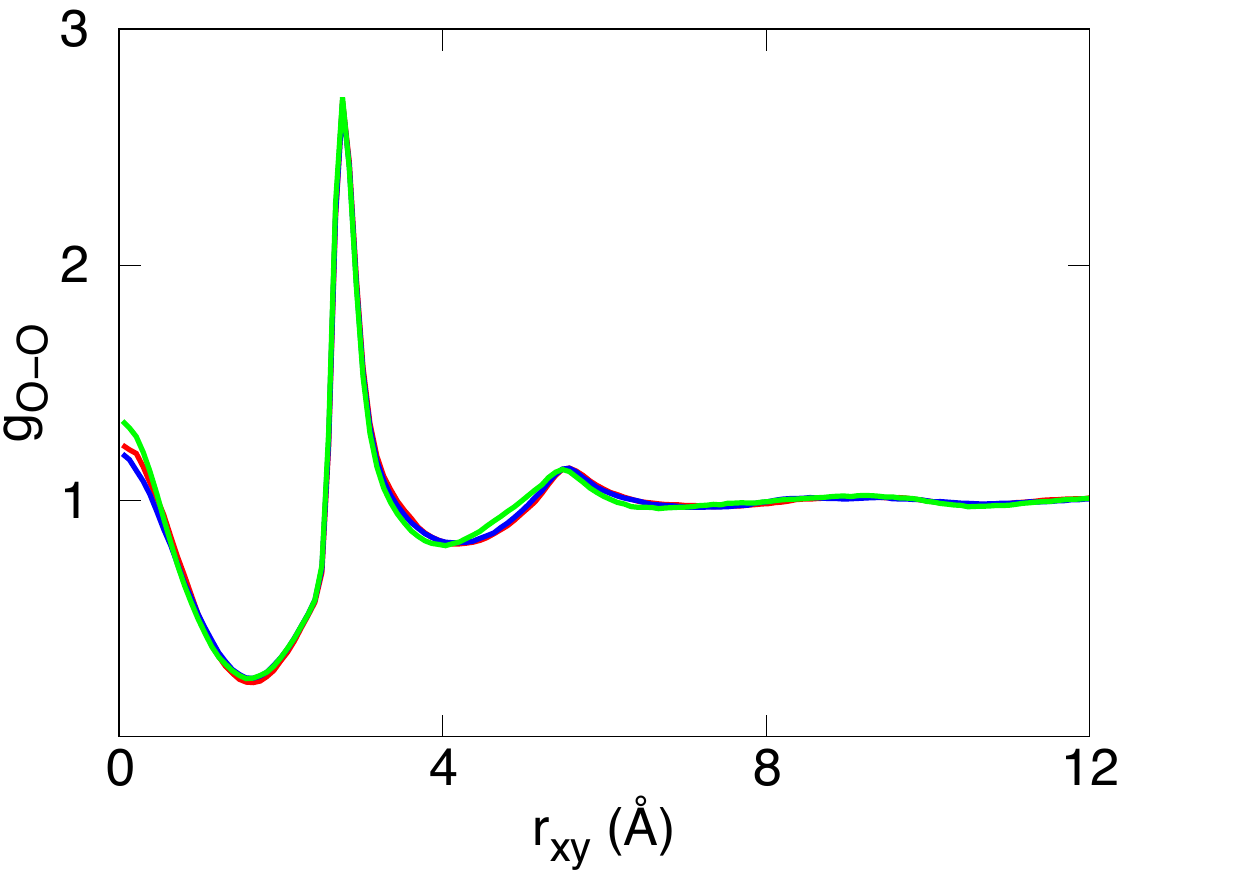}}
\subfigure[\ ]{\includegraphics[width=0.23\textwidth]{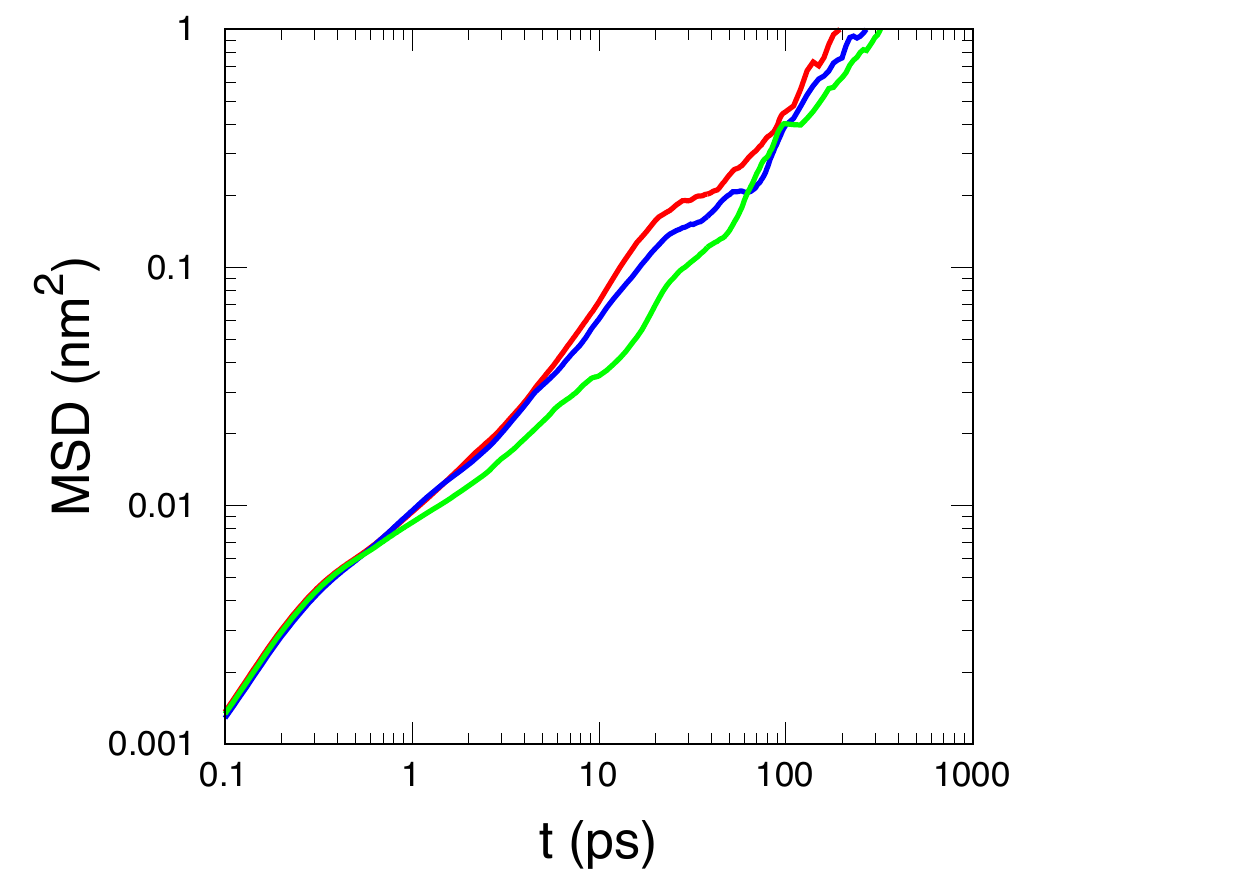}}\\
\subfigure[\ ]{\includegraphics[width=0.23\textwidth]{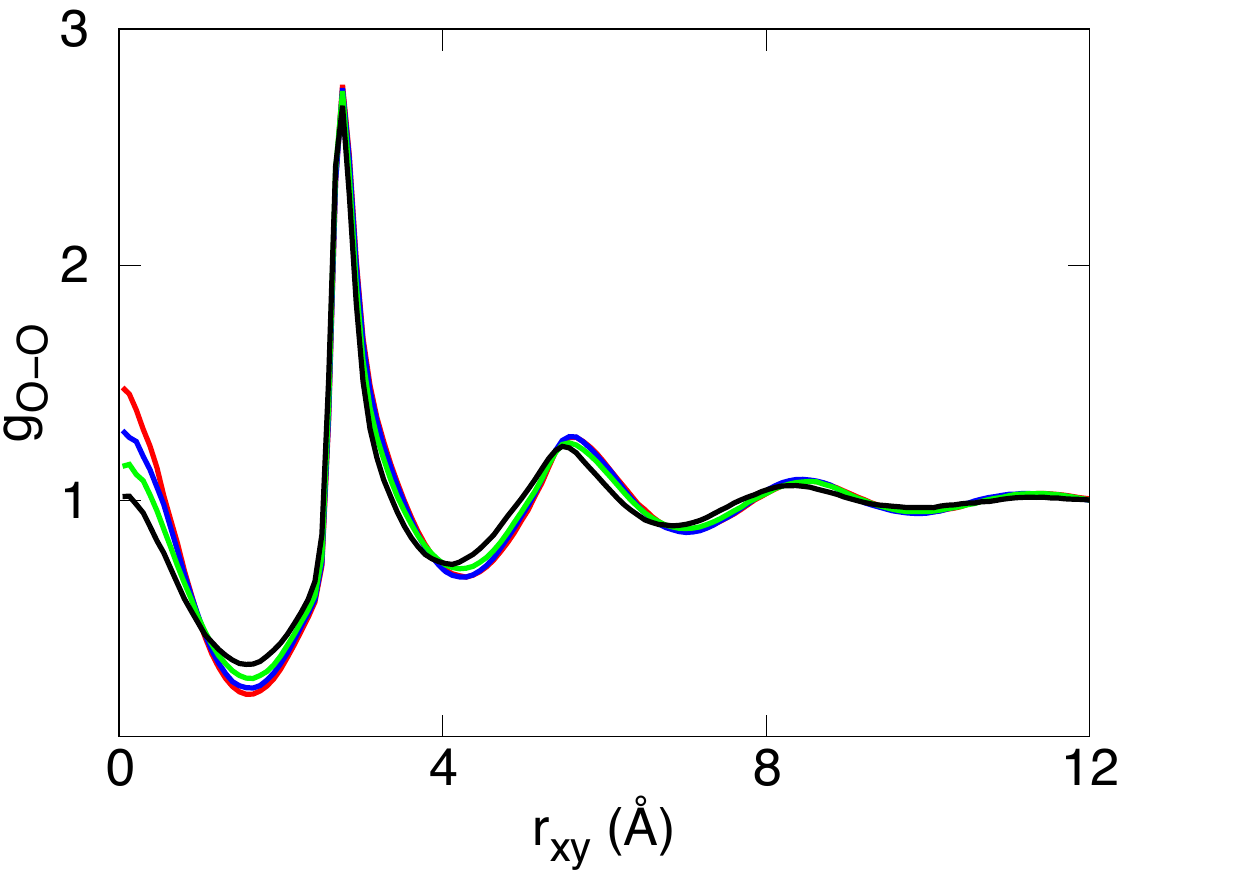}}
\subfigure[\ ]{\includegraphics[width=0.23\textwidth]{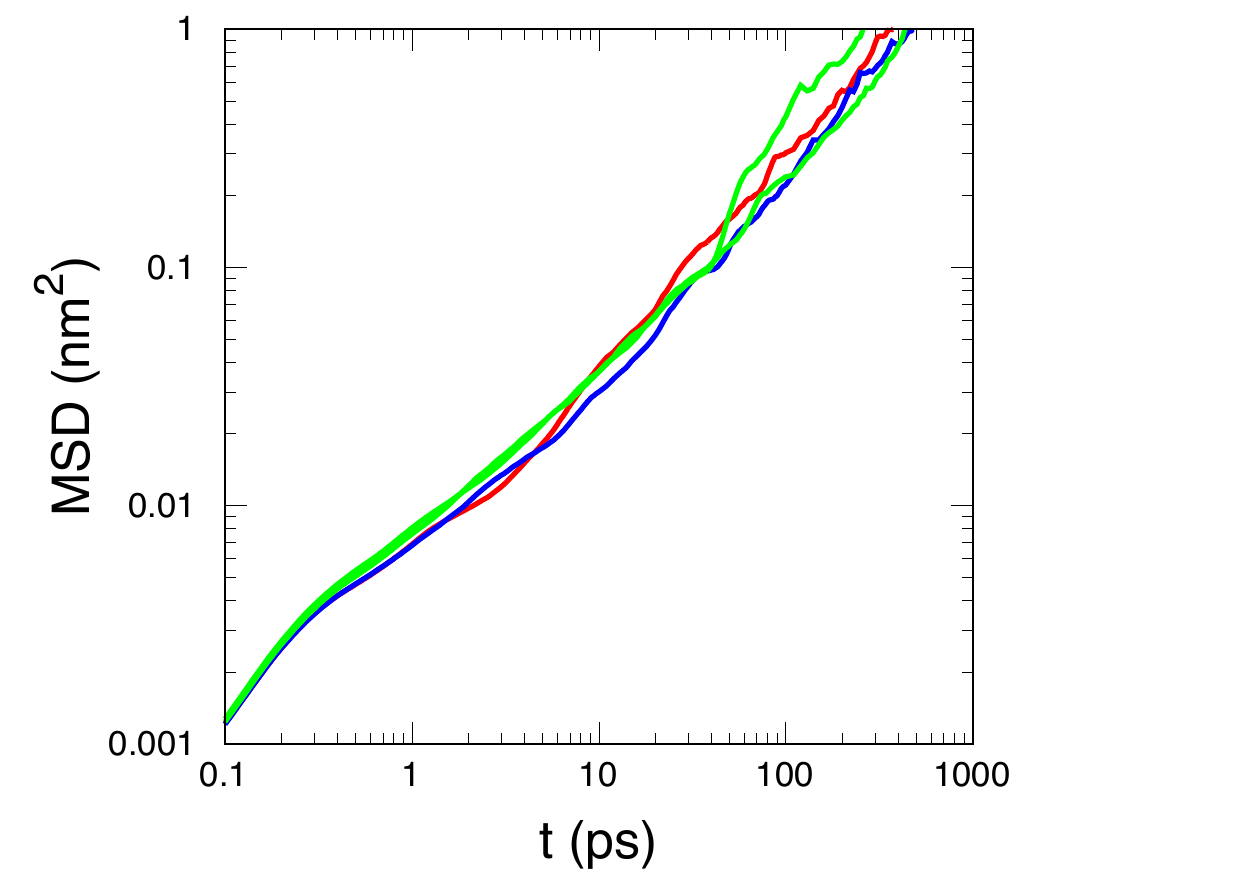}}\\
\caption{Oxygen-oxygen radial distribution function [(a),(c),(e)] and mean-square displacement [(b),(d),(f)] at $\rho$ = 1.17 g~cm$^{-3}$ (a-b), 1.27 g~cm$^{-3}$ (c-d), 1.37 g~cm$^{-3}$ (e-f) for $\alpha$ = 0.0 (red), 0.2 (blue), 0.4 (green), and 0.6 (black), and $a$ = 4.78 $\text{\AA}$.}
\label{fig:liquid3}
\end{figure}

\begin{figure}[t!]
\centering
\subfigure[\ ]{\includegraphics[width=0.21\textwidth]{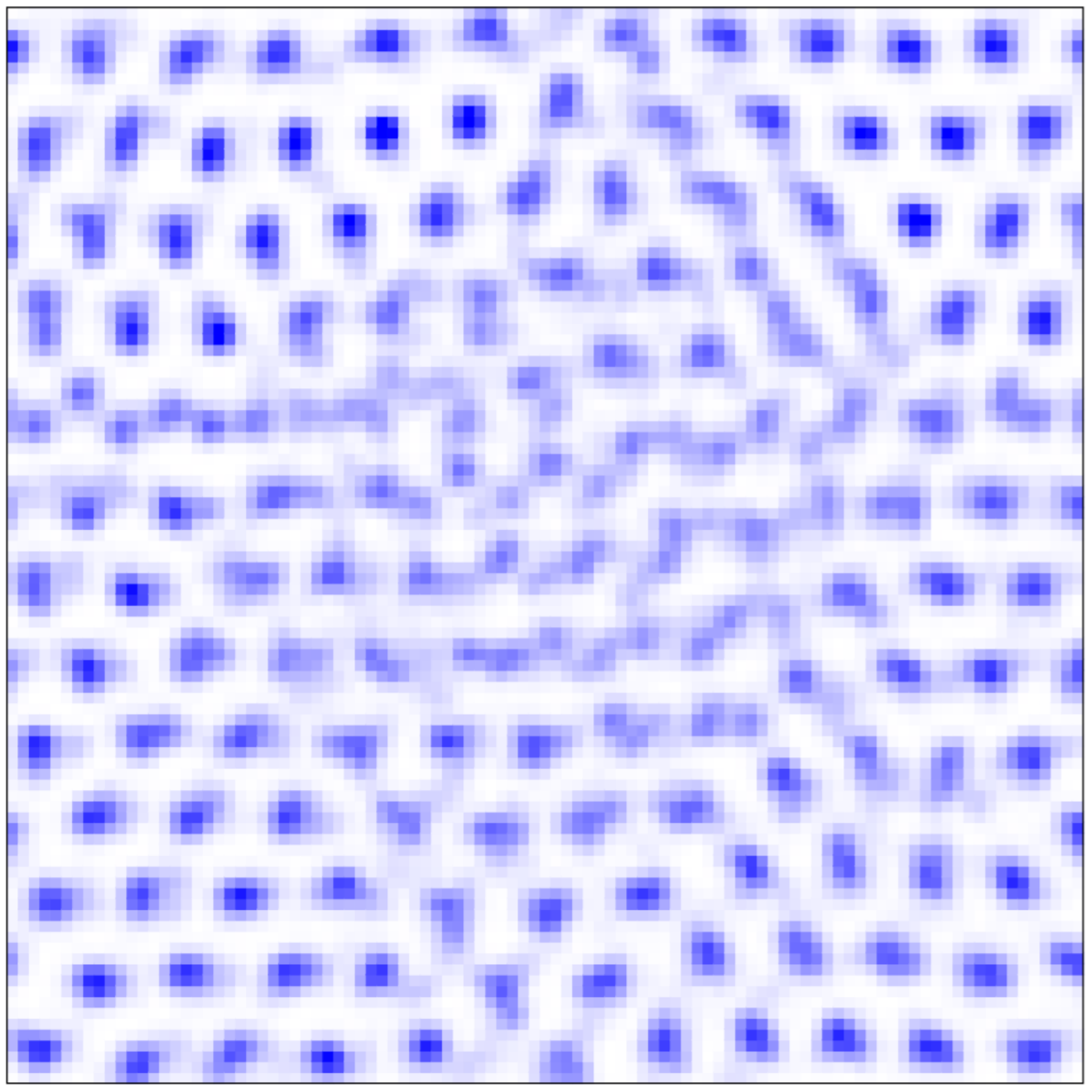}}
\subfigure[\ ]{\includegraphics[width=0.21\textwidth]{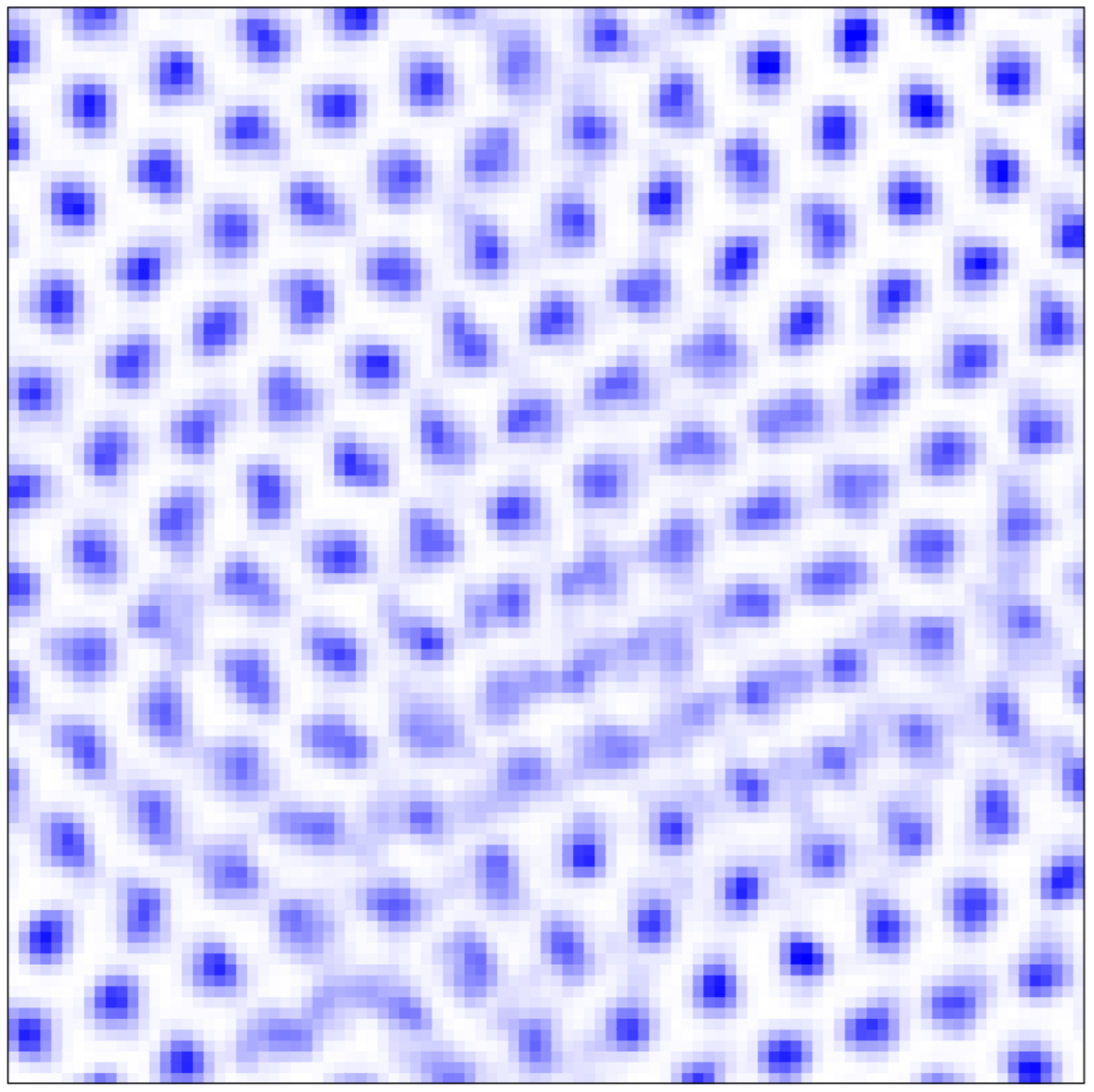}}\\{\hspace{0.38 cm}}
\subfigure[\ ]{\includegraphics[width=0.21\textwidth]{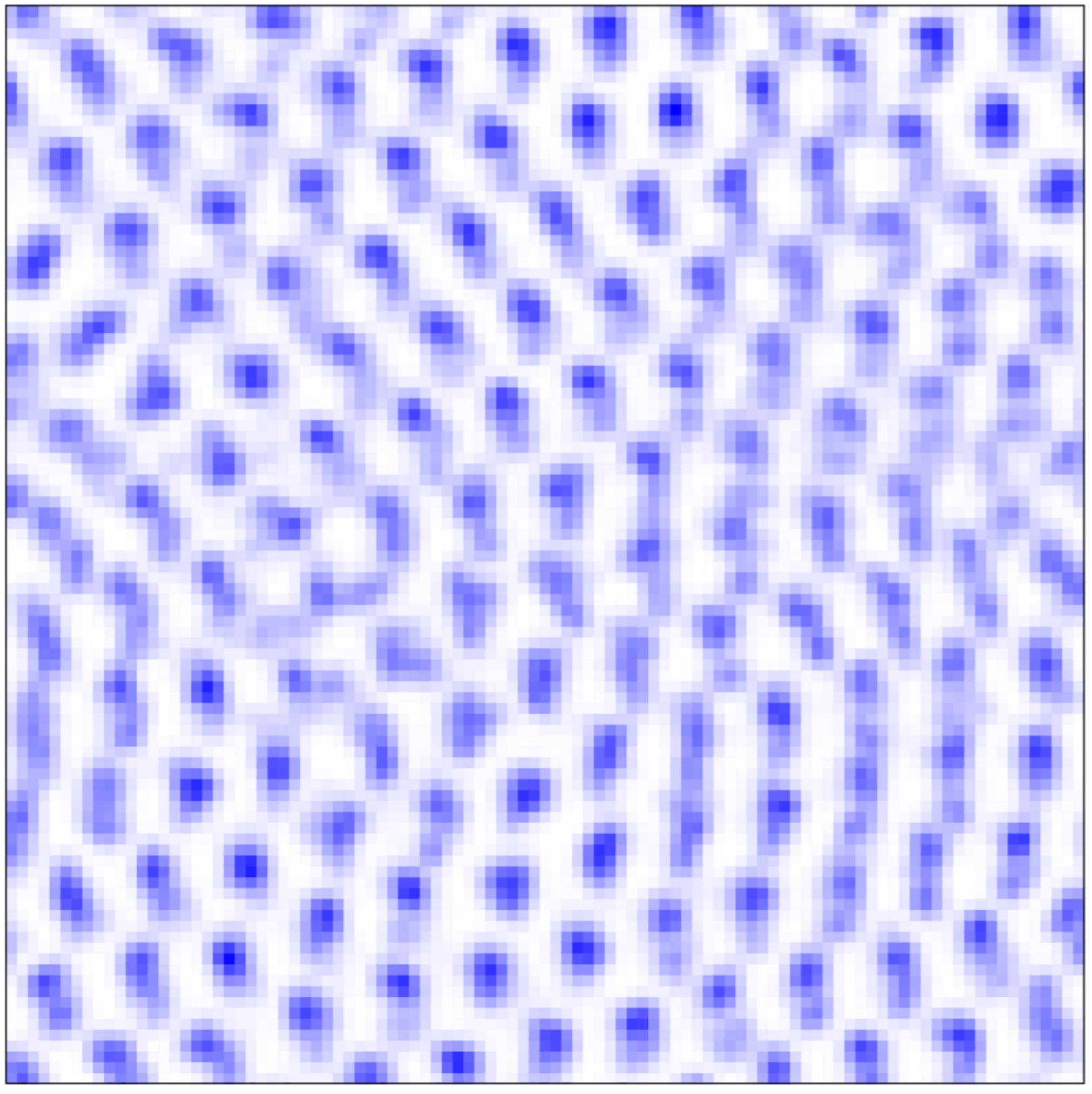}}
\subfigure[\ ]{\includegraphics[width=0.24\textwidth]{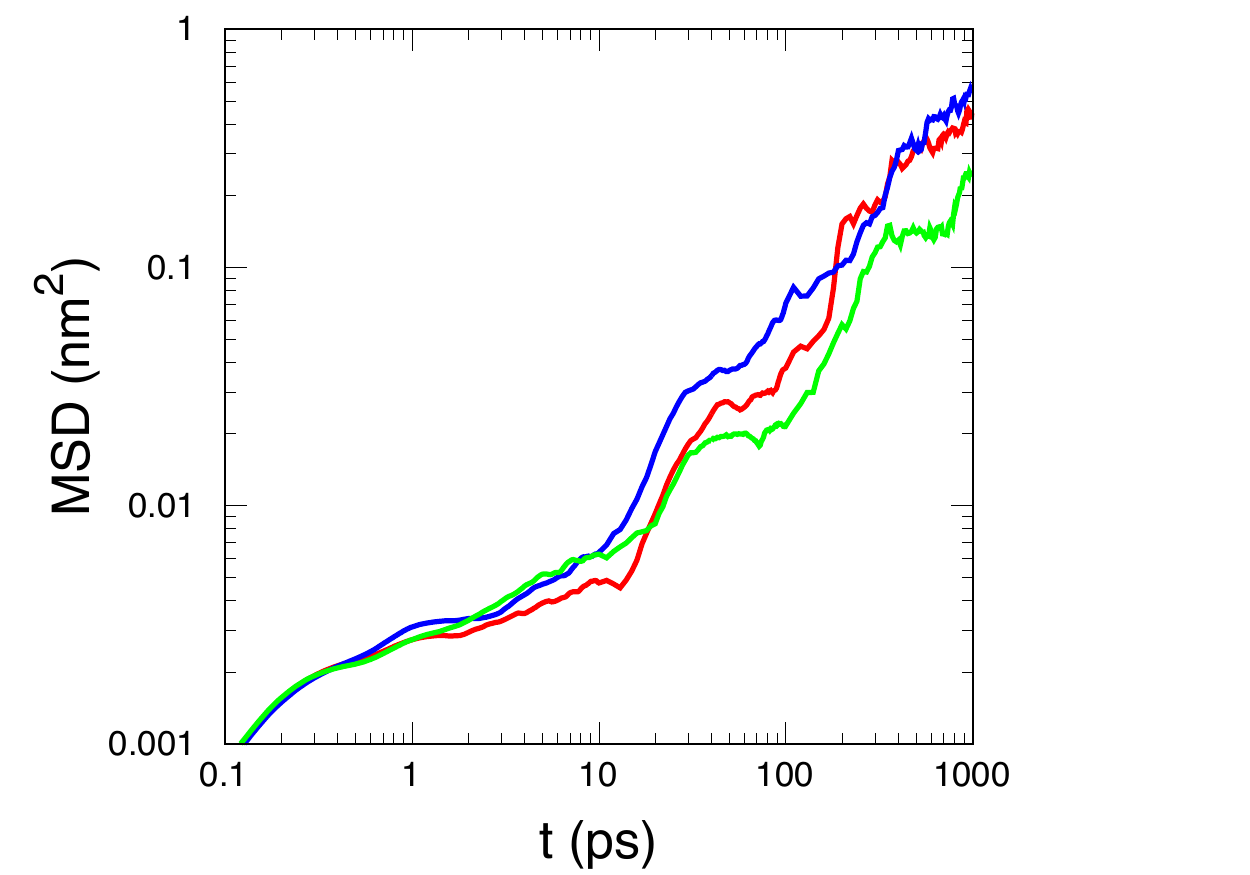}}\\
\caption{$xy$ density histogram of the oxygen atoms at $T$ = 300 K and $\rho$ = 1.47 g~cm$^{-3}$ for $\alpha$ = 1.0 and $a$ = 2.5 $\text{\AA}$ (a), 2.75 $\text{\AA}$ (b), and 3.0 $\text{\AA}$ (c). (d) Mean-square displacement of oxygens at $T$ = 300 K and $\rho$ = 1.47 g~cm$^{-3}$ for $a$ = 2.5 $\text{\AA}$ (red), 2.75 $\text{\AA}$ (blue), and 3.0 $\text{\AA}$ (green).}
\label{fig:hexatic}
\end{figure}

\section{Square tubes ice for $a$ = 4.78 $\text{\AA}$}
\label{A3}
For $a$ = 4.78 $\text{\AA}$, and $0.2\leq\alpha\leq0.6$, we observe the similar high density phases observed in~\cite{zubeltzu2016}. At low temperatures, we observe the formation of the square tubes ice, where both the hydrogen and oxygen atoms are fixed, with the hydrogen atoms tending to point towards the molecules from the same tube. The phase is observed to stay stable for $\alpha\leq$ 0.6. Fig.~\ref{fig:histos} (a) and (b) show the $xy$ density histogram of the oxygen and hydrogen atoms at $T$ = 220 K and $\rho$ = 1.47 g~cm$^{-3}$ for a flat confining potential ($\alpha$ = 0) and for $\alpha$ = 0.6, respectively.

\section{Hexatic phase for $a$ = 4.78 $\text{\AA}$}
\label{A4}

For $a$ = 4.78 $\text{\AA}$ at $\rho$ = 1.47 g~cm$^{-3}$ and $T$ = 300 K corresponding to the reported hexatic phase for $\alpha$ = 0.0~\cite{zubeltzu2016} similar structural features are observed in the RDFs up to $\alpha\leq 0.4$ (Fig.~\ref{fig:rdfhex}) while it clearly changes at $\alpha$ = 0.6.

\section{Honeycomb ice for $a$ = 4.78 $\text{\AA}$}
\label{A5}

We observe that for $a$ = 4.78 $\text{\AA}$ the honeycomb ice stabilizes with $\alpha$ (Fig.~\ref{fig:histices}). When $\alpha$ = 1.0 it is stable at $\rho\leq$ 1.17 g~cm$^{-3}$ for all the sampled temperatures (see Fig.~\ref{fig:PD}).

\section{Liquid for $a$ = 4.78 $\text{\AA}$}
\label{A6}

The liquid shows similar structural and dynamical features before it freezes for $a$ = 4.78 $\text{\AA}$ (Fig.~\ref{fig:liquid3}). The RDFs and MSDs at the different points of the phase diagrams show that for a given $T$ and $\rho$, the diffusivity and structure do not significantly change with $\alpha$, as long as a phase transition does not occur.
{\vspace{0.35 cm}}
\section{Hexatic phase for realistic lattice parameters}
\label{A7}

The hexatic phase shows similar features under the modulation for realistic lattice parameters (Fig.~\ref{fig:hexatic}). The $xy$ density histograms show that the oxygen atoms are arranged into a triangular lattice with the usual shear motion along the main directions of the lattice. The MSDs show similar diffusivities for different realistic lattice parameters ($a$ = 2.50, 2.75, and 3.00 $\text{\AA}$).

\section{Liquid for realistic lattice parameters}
\label{A8}

\begin{figure}[h!]
\centering
\subfigure[\ ]{\includegraphics[width=0.2\textwidth]{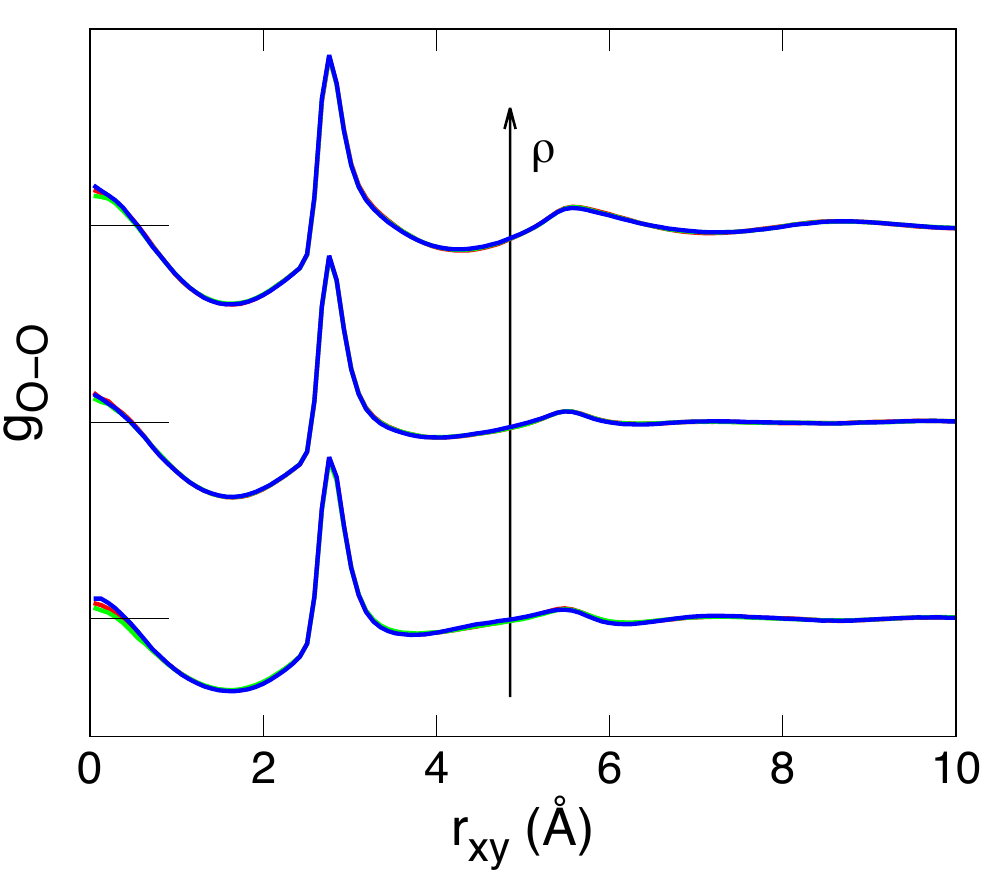}}
\subfigure[\ ]{\includegraphics[width=0.26\textwidth]{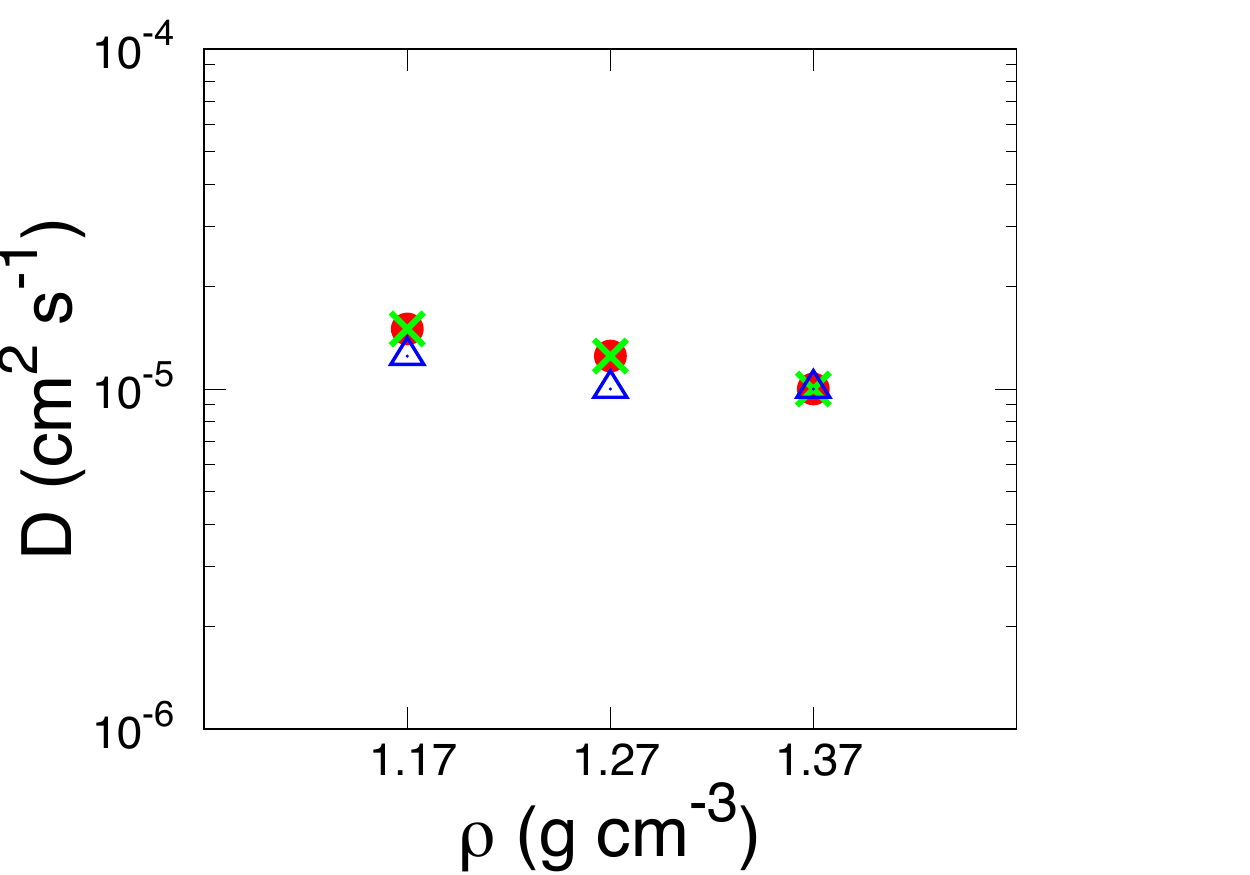}}\\
\subfigure[\ ]{\includegraphics[width=0.2\textwidth]{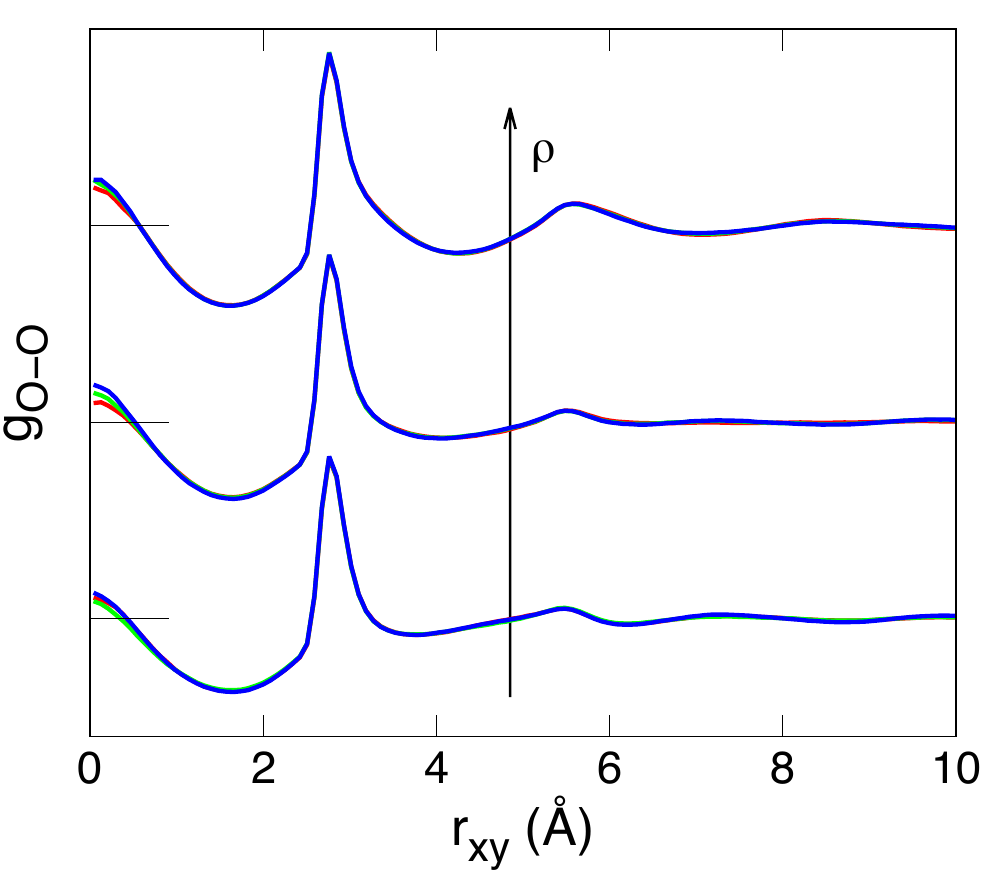}}
\subfigure[\ ]{\includegraphics[width=0.26\textwidth]{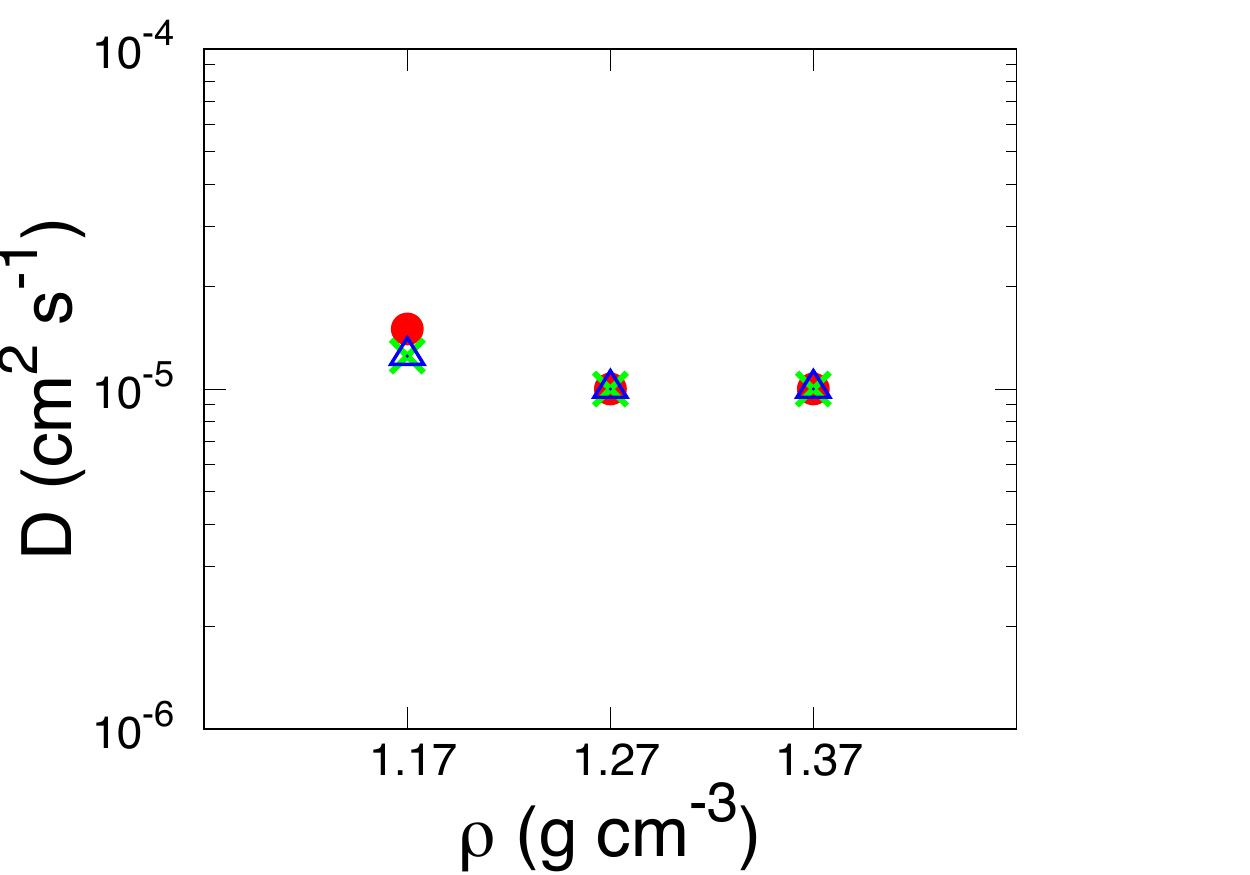}}\\
\subfigure[\ ]{\includegraphics[width=0.2\textwidth]{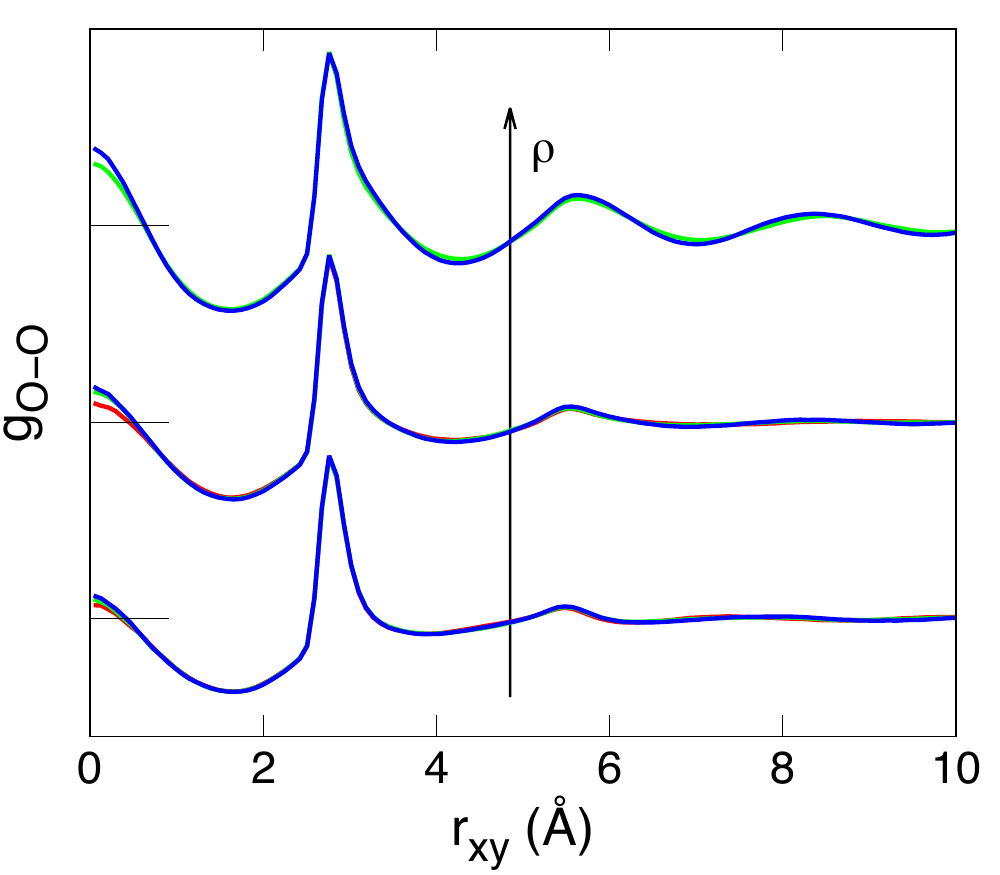}}
\subfigure[\ ]{\includegraphics[width=0.26\textwidth]{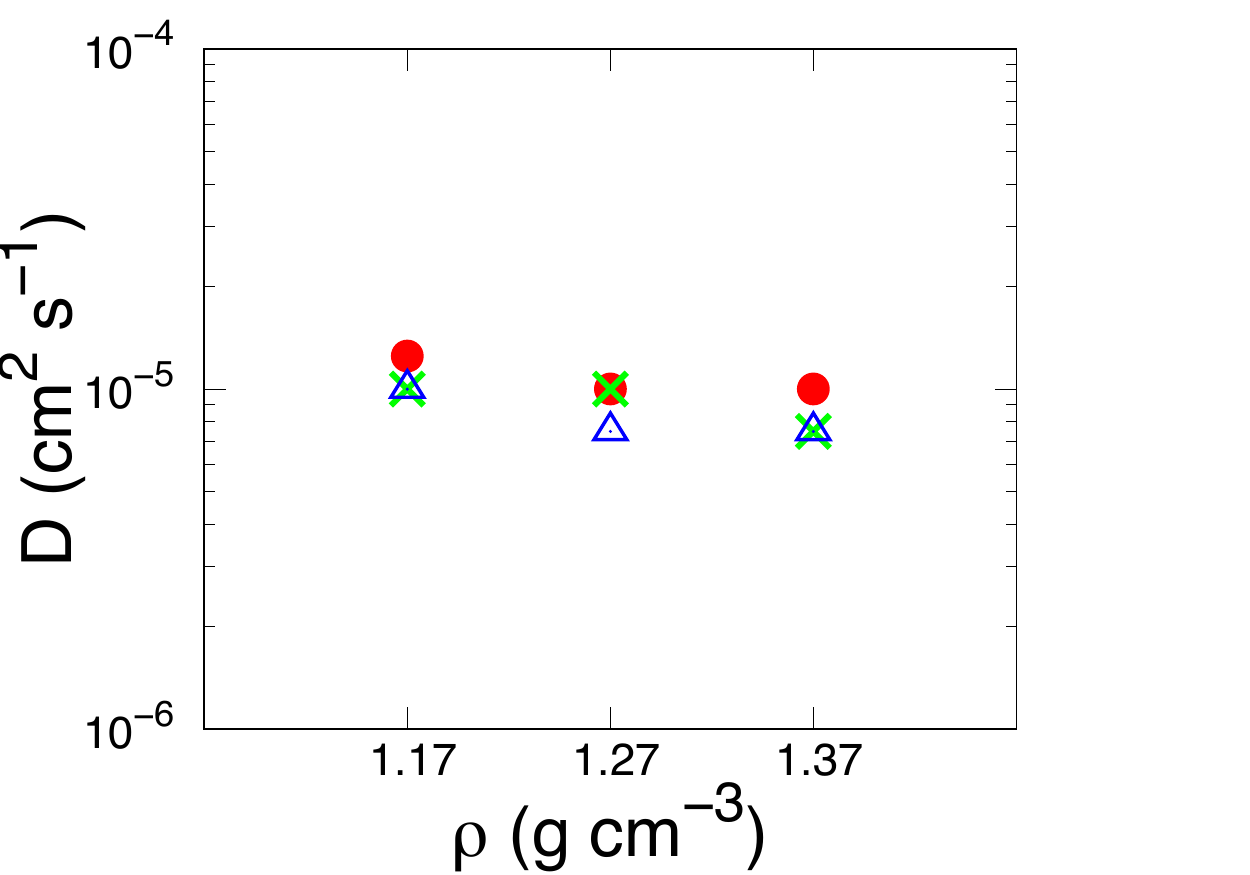}}\\
\caption{Oxygen-oxygen radial distribution function [(a),(c),(e)] and diffusivities [(b),(d),(f)] at $a$ = 2.5 $\text{\AA}$ (a-b), 2.75 $\text{\AA}$ (c-d), and 3.0 $\text{\AA}$ (e-f), at $\rho$ = 1.17 g~cm$^{-3}$ (below), 1.27 g~cm$^{-3}$ (middle), and 1.37 g~cm$^{-3}$ (top) for $\alpha$ = 0.0 (red), 0.2 (blue), 0.4 (green), and 0.6 (black), and $a$ = 4.78 $\text{\AA}$. The curves are shifted on the $y$ axis; the value of saturation of each curve is marked by a horizontal line.}
\label{fig:liquid2}
\end{figure}

The liquid shows similar structural and dynamical features before it freezes for realistic lattice parameters (Fig.~\ref{fig:liquid2}). For the different lattice parameters ($a$ = 2.50, 2.75, and 3.00 $\text{\AA}$) the RDFs are almost indistinguishable at a given $\rho$ and the values of the diffusivity of the liquid are around $10^{-5}$ cm$^2$s$^{-1}$ as for a planar confinement~\cite{zubeltzu2016}.

\end{appendices}

\end{document}